\documentclass[a4paper,11pt]{article}

\usepackage[utf8]{inputenc}
\usepackage{amsmath,amssymb}
\usepackage{hyperref}
\usepackage{graphicx}
\usepackage{epstopdf}
\usepackage{mathrsfs}
\usepackage{mathtools}
\usepackage{float}
\usepackage[normalem]{ulem}  
\usepackage{cite}
\usepackage{subcaption}
\usepackage{titling}
\usepackage{authblk}
\usepackage{setspace}
\usepackage{lipsum}
\usepackage{slashed}
\usepackage{caption}
\usepackage{subcaption}

\makeatletter\@addtoreset{equation}{section}\makeatother

\newcommand{\tr}{\text{tr }}

\newcommand{\preprint}[1]{\begin{table}[t]  
             \begin{flushright}               
             {#1}                             
             \end{flushright}                 
             \end{table}}                     
\renewcommand{\title}[1]{\vbox{\center\LARGE{#1}}\vspace{5mm}}
\renewcommand{\author}[1]{\vbox{\center#1}\vspace{5mm}}
\newcommand{\address}[1]{\vbox{\center\em#1}}
\newcommand{\email}[1]{\vbox{\center\tt#1}\vspace{5mm}}

\newcommand{\ran}{\rangle}
\newcommand{\lan}{\langle}
\newcommand{\f}{\frac}
\newcommand{\nn}{\nonumber}

\def\be{\begin{equation}}
\def\ee{\end{equation}}

\def\l{\lambda}

\def\g{\Gamma}

\def\sl{\slashed}

\def\g{\gamma}
\def\G{\Gamma}

\begin{document}
\begin{titlepage}
\preprint{TIFR/TH/17-38}
\begin{center}
\vskip 1cm
\title{A Complex Fermionic Tensor Model in $d$ Dimensions}
\author{Shiroman Prakash${}^{1,a}$, Ritam Sinha${}^{2,b}$}

\address{${}^1$Department of Physics and Computer Science, Dayalbagh Educational Institute, Dayalbagh, Agra, India 282005}
\address{${}^2$Department of Theoretical Physics, Tata Institute of Fundamental Research, Colaba, Mumbai, India, 400005}

\email{${}^a$shiroman@gmail.com, ${}^b$ritam@theory.tifr.res.in}
\end{center}

\abstract{In this note, we study a melonic tensor model  in $d$ dimensions based on three-index Dirac fermions with a four-fermion interaction. Summing the melonic diagrams at strong coupling allows one to define a formal large-$N$ saddle point in arbitrary $d$ and calculate the spectrum of scalar bilinear singlet operators. For $d=2-\epsilon$ the theory is an infrared fixed point, which we find has a purely real spectrum that we determine numerically for arbitrary $d<2$, and analytically as a power series in $\epsilon$. The theory appears to be weakly interacting when $\epsilon$ is small, suggesting that fermionic tensor models in 1-dimension can be studied in an $\epsilon$ expansion. For $d>2$, the spectrum can still be calculated using the saddle point equations, which may define a formal large-$N$ ultraviolet fixed point analogous to the Gross-Neveu model in $d>2$. For $2<d<6$, we find that the spectrum contains at least one complex scalar eigenvalue (similar to the complex eigenvalue present in the bosonic tensor model recently studied by Giombi, Klebanov and Tarnopolsky) which indicates that the theory is unstable. We also find that the fixed point is weakly-interacting when $d=6$ (or more generally $d=4n+2$) and has a real spectrum for $6<d<6.14$ which we present as a power series in $\epsilon$ in $6+\epsilon$ dimensions.}

\vfill

\end{titlepage}

\eject \tableofcontents

\section{Introduction and Summary}

The Sachdev-Ye-Kitaev (SYK) \cite{Sachdev:1992fk, KitaevTalk} has attracted a great deal of attention recently  \cite{Polchinski:2016xgd, Maldacena:2016hyu, Jevicki:2016bwu, Garcia-Garcia:2016mno, Cotler:2016fpe, Nishinaka:2016nxg, Krishnan:2017ztz, Narayan:2017qtw, Garcia-Garcia:2017pzl, Gross:2017hcz, Das:2017pif, Mandal:2017thl, Gaikwad:2017odv}  as a possibly simple model of holography \cite{Maldacena:1997re, Witten:1998qj, Gubser:1998bc}. 
Tensor models \cite{Gurau:2009tw, Gurau:2011xp, Gurau:2011aq, Gurau:2011xq, Bonzom:2011zz, Tanasa:2011ur, Bonzom:2012hw, Bonzom:2017pqs} have recently been observed  to have dynamics similar to the SYK model \cite{Witten:2016iux, Klebanov:2016xxf, Krishnan:2016bvg}, (but see \cite{Bulycheva:2017ilt, Choudhury:2017tax}).

Several studies of higher-dimensional tensor-models and SYK-like models have been carried out \cite{Turiaci:2017zwd, Murugan:2017eto, Klebanov:2016xxf}. In this note, we consider a tensor model with melonic dominance based on three-index Dirac fermions in $d$ dimensions, with the following action:
\be
S=\int \,d^d x\bigg({i} \bar{\psi}_{a}{}^b{}_{c}\slashed{\partial} \psi^{abc} 
+ 
\f12 g_1 \bar{\psi}_{a_1}{}^{b_1}{}_{c_1}\psi^{a_1 b_2 c_2}
\bar{\psi}_{a_2}{}^{b_1}{}_{c_2}\psi^{a_2 b_2 c_1}
+
\f12 g_2 
\bar{\psi}_{a_1}{}^{b_1}{}_{c_1}\psi^{a_2 b_2 c_1}
\bar{\psi}_{a_2}{}^{b_1}{}_{c_2}\psi^{a_1 b_2 c_2}
\bigg). \label{action}
\ee
The three indices of the fermions transform in the fundamental representation of $U(N)\times O(N)\times U(N)$ (or, more precisely, $SU(N)\times O(N) \times SU(N) \times U(1)$.) This action is a generalization of equation 3.24 of \cite{Klebanov:2016xxf} that contains the most general tetrahedronal interaction\footnote{We thank Igor Klebanov for discussions on this point.} one can write down in $d \leq 2$ dimensions. With a view towards generalizing the one-dimensional theory, it appears natural to set one of the couplings, say $g_2$, to zero, or, instead to set $g_2=-g_1$. We will, however, perform our calculations for arbitrary values of the ratio $g_2/g_1$.

In the large $N$ limit, with $\lambda_i=g_i N^{3/2}$ fixed, the theory is dominated by melonic diagrams, which can be explicitly summed for arbitrary dimension $d$. From dimensional analysis, we know that
\be
[\psi]=\f{(d-1)}2,\hspace{.2cm}[g_i]=2-d
\ee
This implies that in $d<2$, the tetrahedronal coupling is relevant; in $d=2$ the coupling is classically marginal, and in $d>2$, the coupling is irrelevant. 

For $d<2$ (which we treat as continuous) the theory is an infrared fixed point, that (mildly) generalizes the 1-dimensional models studied in \cite{Klebanov:2016xxf}. Based on our results for the spectrum of scaling dimensions of bilinear operators below, we conjecture that this theory is weakly interacting when $\epsilon=2-d$ is small, and could serve as a useful starting point for studying the 1-dimensional theory at finite $N$ in an $\epsilon$-expansion \cite{Wilson:1972cf}.  

The vector model version of the theory \cite{Gross:1974jv}, defined by the strong-coupling limit of the action
\begin{equation}
S_{\text{vector}}= \int d^dx \bigg({i} \bar{\psi}_{i}\slashed{\partial} \psi^{i} + \f12 g \bar{\psi}_{i}\psi^{i}
\bar{\psi}_{j}\psi^{j}\bigg)
\end{equation} 
can be solved in the large $N$ limit in $d>2$ and there is now substantial evidence that this theory has a higher-spin gravitational dual, at least in $d=3$ \cite{Klebanov:2002ja, Sezgin:2002rt, Giombi:2009wh, Giombi:2010vg, Giombi:2011ya} (where it also plays an important role in the bosonization duality \cite{Aharony:2012nh,GurAri:2012is}). In the large $N$ limit, the  vector model can be more rigorously defined as a Legendre transform of the free fermionic theory, by introducing an Hubbard-Stratonovich auxiliary field $\sigma_b$, with the action
\begin{equation}
S= \int d^dx \bigg({i} \bar{\psi}_{i}\slashed{\partial} \psi^{i} + \sigma_b \bar{\psi}_{i}\psi^{i}\bigg).
\end{equation}
This definition is preferable to taking the $g\rightarrow \infty$ limit of an irrelevant $g (\bar{\psi}\psi)(\bar{\psi}\psi)$ interaction term, but from the simple-minded perspective of summing the leading-order diagrams in the large $N$ limit, both approaches give the same results. The UV fixed point can also be studied at finite $N$ in an $\epsilon$ expansion, starting from the Gross-Neveu model in $2+\epsilon$ dimensions, or the Gross-Neveu-Yukawa model in $4-\epsilon$ dimensions \cite{Gross:1974jv,Wilson:1972cf}. (See, e.g., \cite{Ghosh:2015opa,Diab:2016spb,Raju:2015fza,Manashov:2016uam} for recent computations in the vector model.)

For $d>2$, motivated by the vector model case, one might hope that the strong-coupling limit of the melonic theory also formally defines a UV fixed point at large $N$ -- which may have a dual holographic description in $AdS_{d+1}$ that is at least as well-defined as the formal large $N$ solution of the $d$-dimensional bosonic tensor model studied in \cite{Klebanov:2016xxf, Giombi:2017dtl}. One minor advantage of studying the fermionic theory is that the bosonic theory $\phi^4$ rank-three tensor has a direction that is classically unbounded from below --  but this problem is apparently not present for the fermionic tensor model since the fermionic fields are classically Grassmann-valued.

However, in the melonic large-$N$ strong-coupling limit, the scaling dimension of the fermion comes out to be $\frac{d}{4}$. This is below the unitarity bound \cite{Mack:1975je, Minwalla:1997ka} $\frac{d-1}{2}$ for $d>2$, indicating that the fermionic fields cannot be observables in a unitary CFT\footnote{This problem also exists for the tensor models based on higher rank ($q-1$) tensors studied in \cite{Giombi:2017dtl}, where the scaling dimension of the scalar $\Delta_\phi=d/q$, is below the unitary bound if $d>d_*=2(1-\frac{2}{q})^{-1}$. For $q=4$, $d_*=4$ and $q=6$, $d_*=3$.}. To avoid this problem, one might gauge the $SU(N)\times O(N)\times SU(N) \times U(1)$ symmetry so that the individual fermionic fields themselves are not gauge-invariant operators. One of way of doing this in $d=3$ would be using a Chern-Simons field. Assuming that we are able to restrict to the singlet sector consistently, the relevant question is then whether the spectrum of gauge-invariant operators lie above the unitarity bound, which we try to partially address for scalar bilinears in the calculations below. 

\textit{Note Added:} We do not consider the case of $d=2$ in this paper. Shortly after our work appeared, a related paper by Benedetti, Carrozza,  Gurau, and Sfondrini \cite{Benedetti:2017fmp} considers this case in detail and addresses the question of dynamical mass generation.

\subsection{Summary of Results and Discussion}
We first solve for the exact two-point function in the strong coupling limit $\lambda_i \rightarrow \infty$ in Section \ref{two-point}. It is possible to solve the Schwinger-Dyson equations at arbitrary $d$, although the solution is only an IR fixed point for $d<2$. We then consider the strong-coupling limit of the four-point function and solve for the spectrum of spin-$0$ operators, formed from bilinears of the schematic form $\bar{\psi}^{abc}(\ldots)\psi_{a}{}^b{}_c$, closely following \cite{Klebanov:2016xxf} in Section \ref{four-point}. It turns out that the spectrum is essentially independent of the ratio between $\lambda_1$ and $\lambda_2$. Numerical results for the spectrum in various dimensions are presented in Section \ref{numerical}

For $d<2$, no complex eigenvalue is found, and the theory seems well-defined. The scaling dimensions we find suggest that the theory is free in 2 dimensions in the melonic limit, (as one might expect from the Gross-Neveu model), and our analysis allows one to calculate scaling dimensions in $2-\epsilon$ dimensions analytically in a power series in $\epsilon$ in Section \ref{2-epsilon}. It would be interesting to extend our analytic expressions to a finite $N$, which could allow us to study the 1-dimensional fermionic tensor models at finite $N$, starting from the theory in $2-\epsilon$ dimensions. This would require us to study the beta function of the theory at finite $N$ in $2-\epsilon$ dimensions, which we hope to do in the near future. It has been conjectured that the 1-dimensional tensor model is solvable at finite $N$ as well \cite{Krishnan:2017txw}, and it would be interesting to compare results from an $\epsilon$ expansion to an exact or numerical solution.

Though the case $d>2$ may be unphysical, we calculate the spectrum formally in this case as well in Section \ref{numerical}. The theory appears to be weakly interacting when $d=6$ and the spectrum also simplifies drastically in $d=4$.  For $2<d<6$ we find that the spectrum contains a complex eigenvalue similar to the one that is present in the analogous bosonic model \cite{Giombi:2017dtl}, indicating that corresponding fields in a dual gravitational description would lie below the Breitenlohner-Freedman bound.  

In a window $6<d<6.14$, a numerical search suggests that the spectrum contains no complex eigenvalue. Hence, there may be a real fixed point in $6+\epsilon$ dimensions described by this model. The spectrum of bilinear operators appears, however, to contain operators with scaling dimensions below the unitarity bound. We present this spectrum analytically as a power series in $\epsilon$ in Section \ref{6+epsilon}.  Of course, interesting 6-dimensional theories are known to have $N^3$ degrees of freedom \cite{Klebanov:1996un} (and e.g., \cite{Berman:2007bv}), but we do not propose any physical interpretation of this particular theory.

In our calculations, we used dimensional regularization. We also only considered the strong coupling limit of the Schwinger-Dyson equation, \eqref{SD1}, for the two point function. Ideally, one would like to solve the exact Schwinger-Dyson equation carefully, at least numerically, to better understand if this strong-coupling limit is indeed physical. It would also strengthen one's confidence in the existence of the theories in $d>2$ if there was an alternative description as an IR fixed points, similar to the Gross-Neveu-Yukawa model, even if both descriptions have a complex spectrum. 

It is also possible to calculate the spectrum of higher-spin bilinear operators, following \cite{Giombi:2017dtl}. Here there are four different forms for the three-point function $\langle \psi(x_1)\bar{\psi}(x_2)\mathcal O_s(x_3,z)$ (two of which are parity-even and two of which are parity odd), giving rise to four different spectra of spin $s$ operators. It might be interesting to calculate the spectrum, to see if the scaling dimensions are consistent with the unitarity bound, and also to what extent the spectrum is consistent with other general expectations from conformal field theory, e.g., large-spin perturbation theory \cite{Komargodski:2012ek, Alday:2015eya, Kaviraj:2015cxa, Kaviraj:2015xsa, Alday:2016njk, Alday:2016jfr}.

In $d=3$ one can add Chern-Simons gauge fields for any of the symmetry groups. Adding a Chern-Simons field to vector models has been very interesting (e.g., \cite{Giombi:2011kc,  Aharony:2011jz, GurAri:2012is, Aharony:2012nh, Jain:2013py,Jain:2013gza, Bedhotiya:2015uga, Yokoyama:2016sbx}), and affects the spectrum of operators only at the level of $1/N$ corrections (explicitly calculated in \cite{Giombi:2016zwa, MZ}).  Integrating out the gauge field in a tensorial theory would give rise to a ``pillow'' interaction term, with 't Hooft coupling $\lambda_{CS}=\frac{N^2}{k}$. Such an interaction appears to be similar to a large flavor expansion, e.g.,  \cite{Guru:2014ad}, and we expect that this would also only affect $1/N$ corrections to the spectrum we have presented here. Our results may also apply to the large $D$ limit of a $U(N)\times O(D) \times U(N)$ theory, as in \cite{Ferrari:2017ryl, Azeyanagi:2017drg, Azeyanagi:2017mre}.

Of course, the supersymmetric versions of the theory may be more promising, e.g., \cite{Murugan:2017eto}. Perhaps the calculations here may serve as a useful warm-up for a study of these theories.

\section{Two-point function} \label{two-point}
The two-point function of fermions in the free theory is
\be
\lan \psi^{abc}(p)\bar{\psi}_{a'}{}^{b'}{}_{c'}(-q)\ran \equiv G_0(p) \delta^{a}_{a'}\delta^{bb'}\delta^{c}_{c'}\times (2\pi)^d \delta^d(p-q).
\ee
where 
\begin{equation}
G_0(p)=\frac{1}{i\slashed{p}}.
\end{equation}
In the interacting theory, we replace the free propagator $G_0(p)$ with the exact propagator $G(p)$.

\begin{figure}
\begin{center}
\includegraphics[width=10cm]{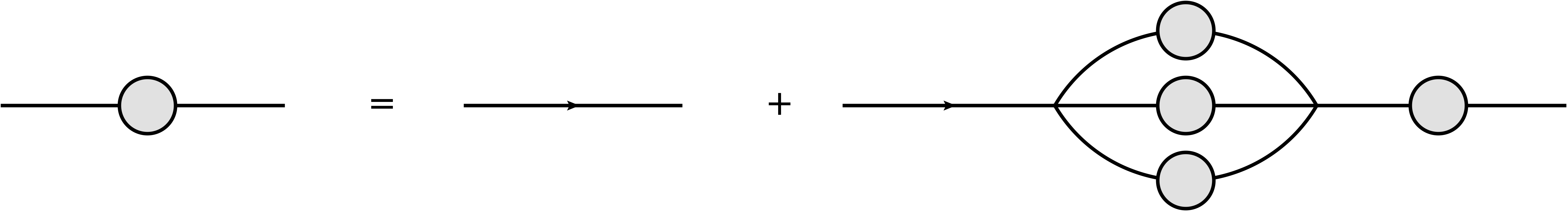}
\caption{The Schwinger-Dyson Equation for the exact propagator. \label{gap}}
\end{center}
\end{figure}

We wish to calculate the two point function in the interacting theory in $d$-dimensions. We sum over all the melonic diagrams in the 
theory in exactly the same way as in, e.g., \cite{Klebanov:2016xxf}. Keeping track of spinor-index contractions, we find the Schwinger Dyson equation, depicted in Figure \ref{gap}, is given by,
\be
\begin{split}
G(p)  = & G_0(p) - (\lambda_1^2+\lambda_2^2) \int \,\f{d^d q \,d^d r}{(2\pi)^{2d}} G_0(p)G(p-q+r)\text{Tr}[G(q)G(r)]G(p) \\
    & +  2 \lambda_1 \lambda_2 \int\, \f{d^d q \,d^d r}{(2\pi)^{2d}} G_0(p)G(p-q+r) G(r) G(q) G(p). 
\end{split}
\ee
This equation can be rewritten as:
\be
\begin{split}
G(p)^{-1} = & G_0(p)^{-1} + (\lambda_1^2+\lambda_2^2) \int \,\f{d^d q \,d^d r}{(2\pi)^{2d}} G(p-q+r)\text{Tr}[G(q)G(r)] \\
	& -  2 \lambda_1 \lambda_2 \int\, \f{d^d q \,d^d r}{(2\pi)^{2d}} )G(p-q+r) G(r) G(q). 
\end{split}
\label{SD1}
\ee

Let us assume $\lambda_2 \sim \lambda_1$, and denote the ratio of $\lambda_2/\lambda_1\equiv\alpha$. We expect the solution to the Schwinger-Dyson equation \eqref{SD1} to be a function of the schematic form 
\begin{equation}
G(p)=f\left(\frac{\lambda_1}{|p|^{2-d}},\alpha\right) \frac{1}{i\slashed{p}}.
\end{equation} 
We are interested in the strong-coupling limit of this solution, which will define a formal large $N$ conformal fixed point. For $d<2$, the strong coupling limit $\lambda_1 \rightarrow \infty$ is equivalent to the IR limit $p \rightarrow 0$; for $d>2$, the strong coupling limit $\lambda_1 \rightarrow \infty$ is equivalent to the UV limit $p \rightarrow \infty$ .  In either limit, we will argue that it is consistent to set $$G(p) \sim \lambda_1^{-1/2} \frac{i \slashed{p}}{p^{d/2+1}}.$$ This implies that $G(p)^{-1} \sim \lambda_1^{1/2} i\slashed{p}(p^2)^{d-2}$. We see that for $d<2$, in the IR limit $|p|\rightarrow 0$, $G^{-1}(p) \gg G_0^{-1}(p)$ and the first term on the RHS of \eqref{SD1} can be dropped. Similarly, for $d>2$, in the UV limit, $|p|\rightarrow \infty$, $G^{-1}(p) \gg G_0^{-1}(p)$ and the first term on the RHS of \eqref{SD1} can be dropped.    

Therefore, to determine the fermion propagator in the strong-coupling limit $\l\rightarrow\infty$, we must solve the equation,
\be
\begin{split}
G(p)^{-1} = & (\lambda_1^2+\lambda_2^2) \int \,\f{d^d q \,d^d r}{(2\pi)^{2d}} G(p-q+r)\text{Tr}[G(q)G(r)] \\
	& -  2 \lambda_1 \lambda_2 \int\, \f{d^d q \,d^d r}{(2\pi)^{2d}} )G(p-q+r) G(r) G(q). 
\end{split}
\label{SD}
\ee
Our aim is to find the solution to the above self-consistent equation. We assume the following general ansatz for the solution,
\be
G(p)=A(\lambda_1,\lambda_2) \f{i\slashed{p}}{(p^2)^{\alpha}}
\ee

Substituting this ansatz into equation \eqref{SD} and carefully performing the integrals, to determine the numerical factors $A$ and $\alpha$, we find the exact propagator is given by,
\be
G(p)=-\l^{-1/2}\f{i\sl{p}}{(p^2)^{d/4+1/2}}\bigg[\f{d_{\g}}{(4\pi)^d}\f{\G(1/2 - d/4)}{\G(3d/4 +1/2)}\bigg]^{-1/4}. \label{two-point-momentum-space}
\ee
where 
\be
\lambda^2=(\lambda_1^2+\lambda_2^2) -2\lambda_1\lambda_2/d_\gamma.
\ee
and $d_\gamma$ denotes the dimensionality of the Dirac gamma-matrices in $d$-dimensions. 

Translating to position space, we obtain:
 \begin{align}
  G(x)&=\f1{(2\pi)^d}\int \,d^d p e^{-i p\cdot x} G(p) = i \g^{\mu} A \f1{(2\pi)^d}\int \,d^d p e^{-i p\cdot x}\f{p_{\mu}}{(p^2)^{d/4+1/2}}\nn\\
  &=-\l^{-1/2}\bigg[\f1{d_{\g}\pi^{d}}\f{\G(3d/4+1/2)}{\G(1/2-d/4)}\bigg]^{1/4}\f{\sl{x}}{(x^2)^{d/4+1/2}} 
 \end{align}
When we reduce to $d=1$, this solution agrees with Equation 3.11 of \cite{Klebanov:2016xxf}. Note that the gap equation has been solved numerically for arbitrary $\lambda$ in $d=1$ in \cite{Maldacena:2016hyu}, which helps determine the correct sign of $A$.

We see that the scaling dimension of the fermionic field in the strong-coupling fixed-point can be taken to be $\Delta_\psi = \frac{d}{4}$. This immediately raises a concern that the scaling dimension of the fermion will be below the unitarity bound ($(d-1)/2$ for $d>2$, which suggests that the theories we study do not exist for $d>2$. However, as mentioned in the introduction, if we gauge the $SU(N)\times O(N) \times SU(N) \times U(1)$ symmetry (say with a Chern-Simons field in $3$ dimensions), to restrict to the singlet sector, then the fermionic operators themselves would not be gauge invariant, (and would probably not have a well defined scaling dimension at order $1/N$, if the strength of the Chern-Simons gauge field were non-zero.) In that case, one should only check if scaling dimensions of gauge-invariant operators, such as the bilinears we study below, have scaling dimensions above the unitarity bound.

\section{Four-point Function and Spectrum}
\label{four-point}

In this section, we will set-up the necessary ingredients to obtain the spectrum of spin-0 bilinears. We will closely follow the now standard method used by, e.g., \cite{Klebanov:2016xxf} in $d$-dimensions. (We remark that one should question to what extent the strong coupling limit is rigorous in higher dimensions, particularly without a numerical solution for intermediate values of $\lambda$.)

\begin{figure}
\begin{center}
\includegraphics[width=8cm]{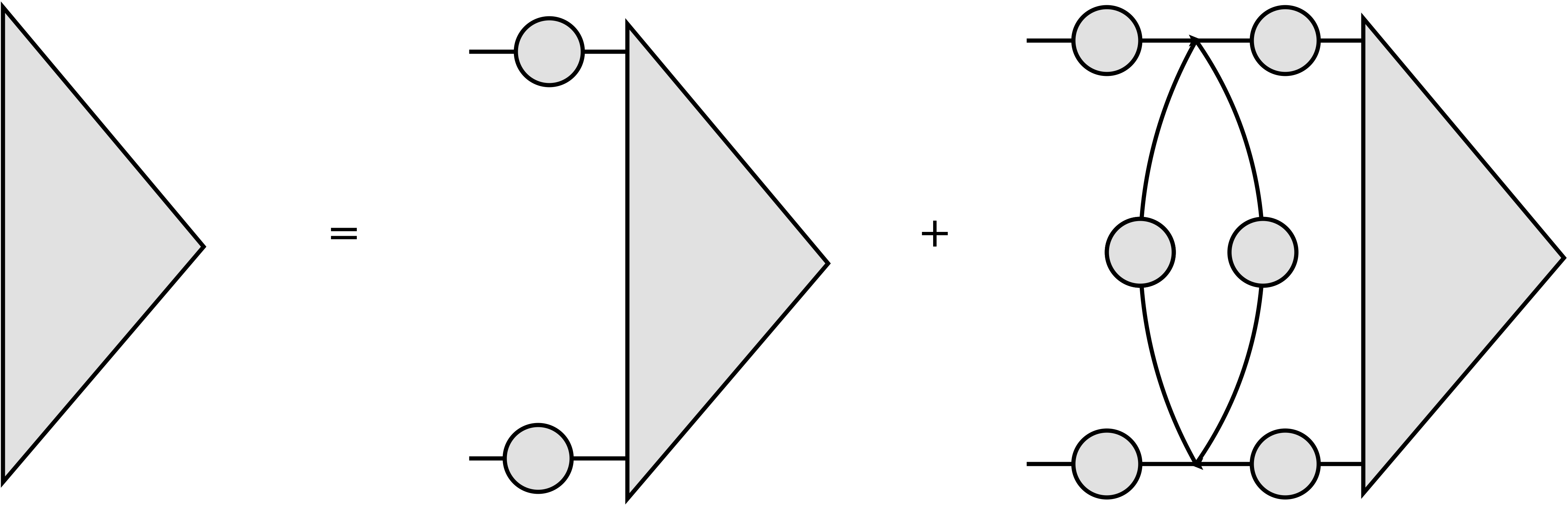}
\caption{The Schwinger-Dyson Equation for the exact three point function $\langle \psi(x_1) O_s(x_3,\epsilon_3) \bar{\psi}(x_2)\rangle$. \label{kernel-picture}}
\end{center}
\end{figure}

The essential idea is that the exact three-point function of a bilinear operator with two fermionic fields $\langle \psi(x_1)  \bar{\psi}(x_2) \mathcal O_s(x_3)\rangle$ obeys a Schwinger-Dyson equation depicted schematically in Figure \ref{kernel-picture} above. In the strong coupling limit, we can drop the first term on the RHS of this equation. This implies that, in the conformal fixed point, $\langle \psi(x_1) \bar{\psi}(x_2) \mathcal O_s(x_3)\rangle$  must be an eigenvector of the ``integration kernel" depicted in the second term of the RHS with eigenvalue 1. Solving for the eigenvectors of the integration kernel determines the allowed forms of the three-point function, which in turn determines the allowed scaling dimensions of operators $\mathcal O_s$ in the fixed point. 

\subsection{Bilinear Operators}
There are various bilinear operators whose scaling dimensions we would like to calculate. 

The spin-0 bilinears are operators of the schematic form $\bar{\psi}(\slashed{\partial})^n \psi$. Note that, for odd $n$, this is a parity-even scalar in $d=3$, while for even $n=2m$, this can be written as $\bar{\psi}(\partial^2)^m \psi$, which is a parity-odd pseudo-scalar in $d=3$. 

There are also higher spin operators, which generalize the free currents explicitly given in \cite{Giombi:2011kc} (see also \cite{Giombi:2016zwa}) to include extra derivatives. For spin 1, these could take the form $\bar{\psi}\gamma_\mu (\slashed{\partial})^n   \psi$, which, in $d=3$, is a vector for $n$ even, and a pseudo-vector for $n$ odd. Another form that these might take is $\bar{\psi}\overleftrightarrow{\partial}_\mu (\slashed{\partial})^n \psi$, which, in $d=3$, is a pseudo-vector for $n$ even, and a vector for $n$ odd. In higher dimensions, there are also operators with anti-symmetric indices (e.g., $\bar{\psi}\gamma_{\mu\nu}\psi$) and mixed symmetry indices. In the present work, we do not consider these operators.

\subsection{Allowed Forms for the Three-Point Function}
We will restrict our attention to the parity-even and parity-odd scalar operators, as these are the operators of physical significance in $d<2$ dimensions. However, it would be possible to calculate the spin $s$ spectrum for the theory using calculations similar to what we present here. 

Let us consider the three-point function:
\begin{equation}
\langle \psi(x_1) \bar{\psi}(x_2) \mathcal O(x_3) \rangle.
\end{equation}
where $\mathcal O$ is a bilinear operator of spin $0$ and scaling dimension $\tau$. Let us temporarily restrict our attention to $d=3$, where we have seen that the operator may be either parity-even or parity-odd. As in \cite{Klebanov:2016xxf}, we will use the allowed forms of this three-point function to derive eigenvectors of the integration kernel described below. 

The most general three point function $\langle\psi(x_1)  \bar{\psi}(x_2) O(x_3) \rangle$ including both parity-even and odd contributions can be written as \cite{GPY, Costa:2011mg, Osborn:1993cr}:
\begin{equation}
\langle (\bar{\xi}_1\psi(x_1))  (\bar{\psi}(x_2)\xi_2) {\mathcal O}_s(x_3) \rangle = \frac{ a P_3 + b  (S_3/P_3)}{|x_{31}|^\tau |x_{12}|^{2\Delta_\psi-1 - \tau} |x_{23}|^\tau } 
\end{equation} 

For general $x_3$ these forms can be written as:
\begin{eqnarray}
P_3 & \sim & \bar{\xi}_1 {\slashed{\breve{x}}_{12}} \xi_2 \\
 (S_3/P_3) & \sim & \frac{\bar{\xi}_1 \slashed{x}_{13} \slashed{x}_{32}\xi_2}{|x_{12}||x_{31}||x_{23}|} 
\end{eqnarray}
where we define $\breve{x}^\mu=\frac{x^\mu}{x^2}$. 

It is convenient to eliminate $x_3$, since the integration kernel derived below does not involve $x_3$. In the limit $|x_3| \rightarrow \infty$ we find:
\begin{eqnarray}
P_3 & \sim & \frac{\slashed{x}_{12}}{|x_{12}|^2}\\
(S_3/P_3) \Big|_{1 \leftrightarrow 2} & \sim & \frac{\mathbf{1}}{|x_{12}|}
\end{eqnarray}
In the above, we may have dropped some numerical factors relative to the definitions in \cite{GPY}. We also removed the polarization spinor $\bar{\xi}_1 = \xi_1^\alpha \epsilon_{\alpha\beta}$ from the left and $\xi_2$ from the right.  


From these expressions, we see the ansatz for eigenvectors of the integration kernel corresponding to parity-even scalar operators is
\begin{equation}
v^{\text{even}}_{d,\tau}(x_1,x_2) = a \frac{\slashed{x}_{12}}{|x_{12}|^{d/2+1 - \tau}} \label{parity-even-fermionic-eigenvectors},
\end{equation}
and the the ansatz for eigenvectors corresponding to parity-odd scalar operators is
\begin{equation}
v^{\text{odd}}_{d,\tau}(x_1,x_2) = b\frac{1}{|x_{12}|^{d/2- \tau}}  \label{parity-odd-fermionic-eigenvectors}.
\end{equation}
Although our derivation above assumed $d=3$, we expect that this ansatz is valid in $d$ dimensions. We also used $\Delta_\psi= d/4$ in the above expressions.


\subsection{Integration Kernel}

To write down the integration kernel in a simple form (i.e., without many free spinor indices), it is convenient to denote the bilinear operator whose three-point function we are calculating as $$\mathcal O = \bar{\psi}^{a_2b_2c_2}(x_3)\mathcal V \psi^{a_2b_2c_2}(x_4) \Big|_{x_3=x_4=x}.$$ The operator $\mathcal V$ could be proportional to a Dirac matrix, $\gamma_\mu$, or the identity, $\mathbf 1$, and may involve derivative operators as well.

\begin{figure}[t]
\centering
\begin{subfigure}{.5\textwidth}
  \centering
  \includegraphics[width=.8\linewidth]{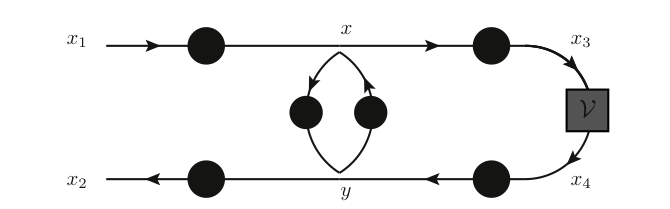} \includegraphics[width=.8\linewidth]{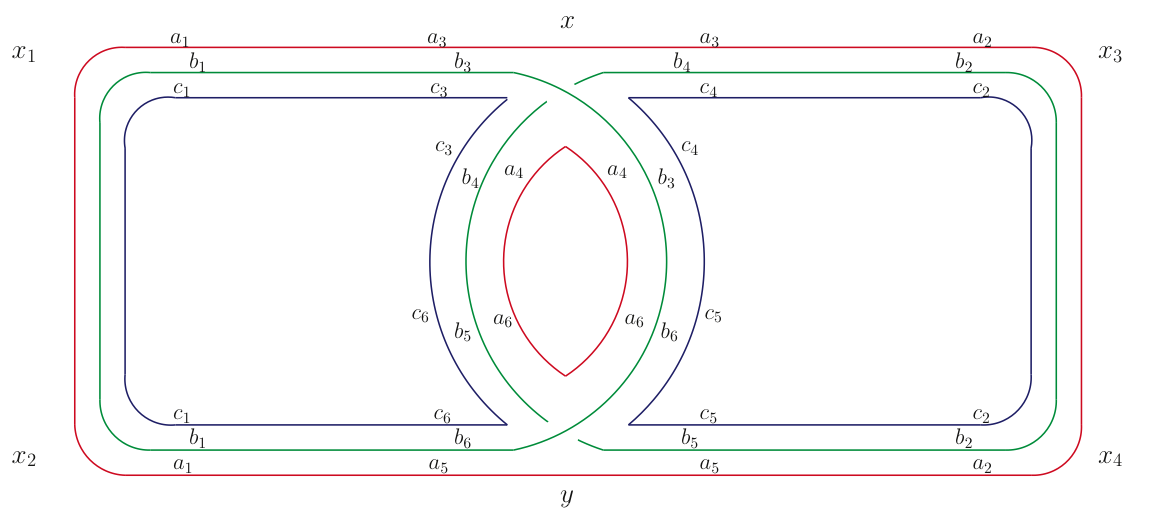} 
  \caption{}
\end{subfigure}%
\begin{subfigure}{.5\textwidth}
  \centering
  \includegraphics[width=.8\linewidth]{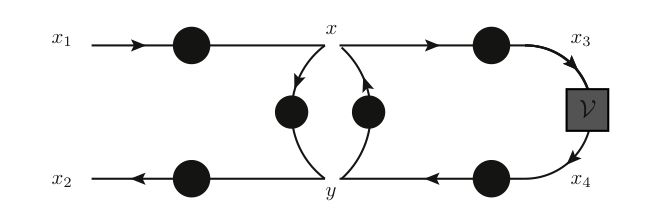} \includegraphics[width=.8\linewidth]{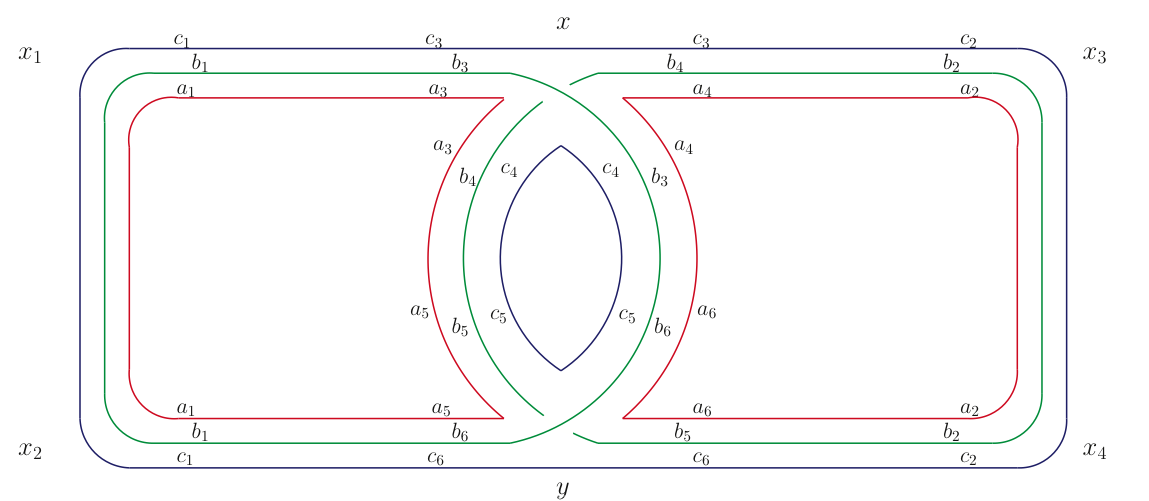} 
  \caption{}
\end{subfigure}
\begin{subfigure}{.5\textwidth}
  \centering
  \includegraphics[width=.8\linewidth]{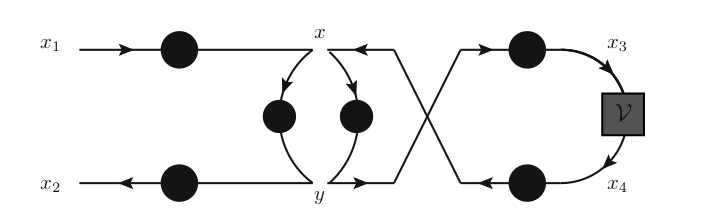} \includegraphics[width=.8\linewidth]{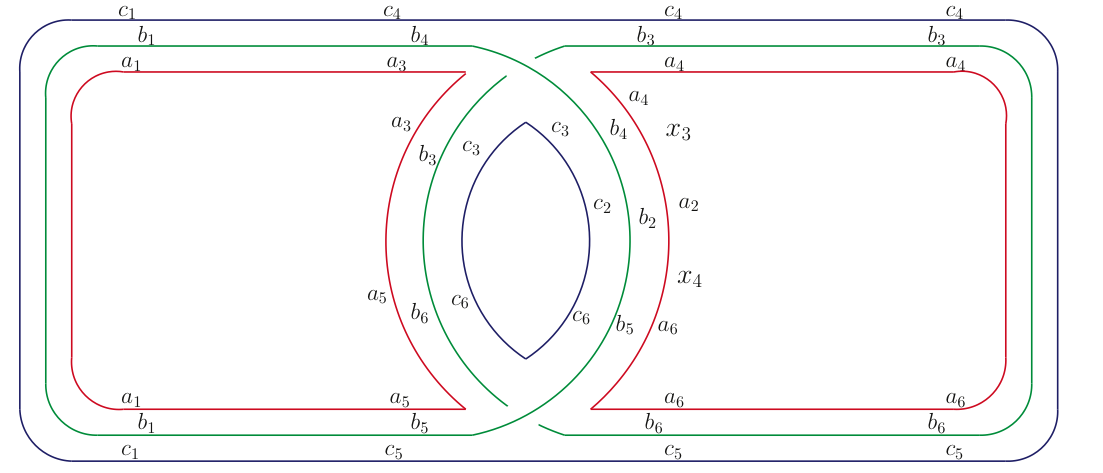} 
  \caption{}
\end{subfigure}
\caption{Feynman diagrams corresponding to melonic  Wick contractions proportional to $\lambda_1^2$ or $\lambda_2^2$. The above diagram shows contraction of spinor indices, and the lower diagram shows contraction of colour indices.}
\label{fig3}
\end{figure}

\begin{figure}[t]
\centering
\begin{subfigure}{.5\textwidth}
  \centering
  \includegraphics[width=.8\linewidth]{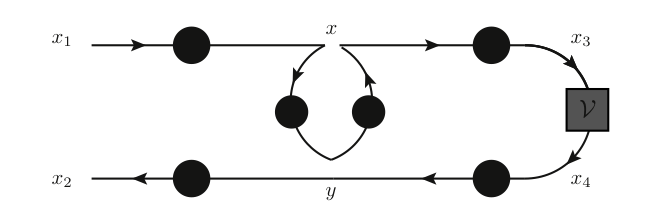} \includegraphics[width=.8\linewidth]{fig_2c_4c} 
  \caption{}
\end{subfigure}%
\begin{subfigure}{.5\textwidth}
  \centering

  \includegraphics[width=.8\linewidth]{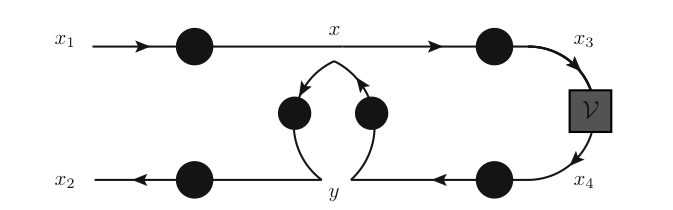}  \includegraphics[width=.8\linewidth]{fig_1c_5c} 

  \caption{}
\end{subfigure}
\begin{subfigure}{.5\textwidth}
  \centering
  \includegraphics[width=.8\linewidth]{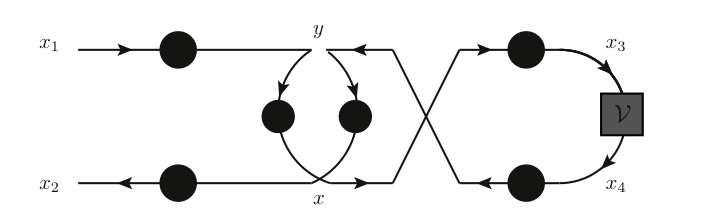}  \includegraphics[width=.8\linewidth]{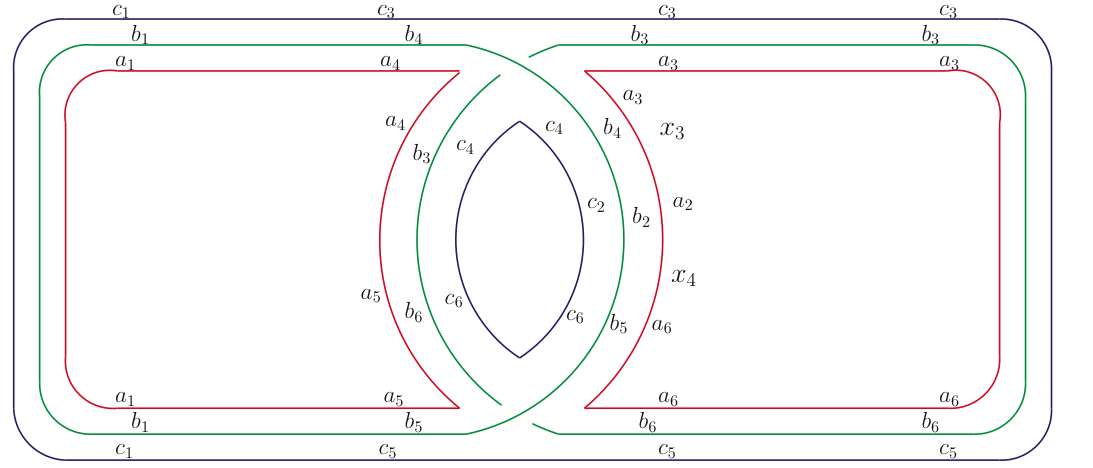} 
  \caption{}
\end{subfigure}

\caption{Feynman diagrams corresponding to melonic Wick contractions proportional to $\lambda_1 \lambda_2$. The above diagram shows contraction of spinor indices, and the lower diagram shows contraction of colour indices.}
\label{fig4}
\end{figure}

To evaluate the fermionic kernel in the large $N$ limit, we need to consider all the melonic Wick contractions of 
\begin{equation}
\begin{split}
& \langle \psi^{a_1 b_1 c_1}(x_1) \left(\bar{\psi}^{a_2b_2c_2}(x_3)\mathcal V \psi^{a_2b_2c_2}(x_4) \right) \bar{\psi}^{a_1 b_1 c_1}(x_2)  \rangle\Big|_{g^2} = \\
&  \frac{1}{2! 2^2} \psi^{a_1 b_1 c_1}(x_1) \left(\bar{\psi}^{a_2b_2c_2}(x_3)\mathcal V \psi^{a_2b_2c_2}(x_4) \right) \\
& 
\int d^dx ~ \Big[ g_1 (\bar{\psi}^{a_3 b_3 c_3}(x) {\psi}^{a_3 b_4 c_4}(x))(\bar{\psi}^{a_4 b_3 c_4}(x) {\psi}^{a_4 b_4 c_3}(x)) \\ 
& + g_2 (\bar{\psi}^{a_3 b_3 c_3}(x) {\psi}^{a_4 b_4 c_3}(x))(\bar{\psi}^{a_4 b_3 c_4}(x) {\psi}^{a_3 b_4 c_4}(x)) \Big]\\
& 
\int d^dy ~\Big[ g_1(\bar{\psi}^{a_5 b_5 c_5}(y) {\psi}^{a_5 b_6 c_6}(y))(\bar{\psi}^{a_6 b_5 c_6}(y) {\psi}^{a_6 b_6 c_5}(y)) \\
& + g_2 (\bar{\psi}^{a_5 b_5 c_5}(y) {\psi}^{a_6 b_6 c_5}(y))(\bar{\psi}^{a_6 b_5 c_6}(y) {\psi}^{a_5 b_6 c_6}(y)) \Big] \bar{\psi}^{a_1 b_1 c_1}(x_2).
\end{split}
\end{equation}
These are pictured in Figures \ref{fig3} and \ref{fig4}.


Let us define the zeroth order ladder diagram as $\Gamma_0 = G(x,x_3) \mathcal V G(x_4,y) \equiv v(x,y)$. Processing this expression, the fermionic integration kernel can be found to be:
 \begin{equation}
 \begin{split} K[v(x,y);x_1,x_2] =  \int d^dx d^dy \Big( & -(\lambda_1^2+\lambda_2^2) G(x_1,x)v(x,y)G(y,x_2) \tr [G(x,y)G(y,x)]  \\
 							& -(\lambda_1^2+\lambda_2^2) G(x_1,x)G(x,y)G(y,x_2) \tr [G(y,x) v(x,y)] \\
 							&  -(\lambda_1^2+\lambda_2^2) G(x_1,y)G(y,x)G(x,x_2) \tr[G(y,x)v(x,y)]  \\
 							& + 2\lambda_1 \lambda_2 G(x_1,x)G(x,y)G(y,x)v(x,y)G(y,x_2) \\
 							& + 2\lambda_1 \lambda_2 G(x_1,x)v(x,y)G(y,x)G(x,y)G(y,x_2) \\
 							& + 2\lambda_1 \lambda_2 G(x_1,y)G(y,x)v(x,y)G(y,x)G(x,x_2) \Big)
\end{split}  
\end{equation}
The strong-coupling limit of the exact three-point functions $\langle \psi(x)\bar{\psi}(y)O_s(x_3,\epsilon)\rangle \equiv v_{\tau,s}$, whose general forms were given in the previous section, must be eigenvectors of this integration kernel,
\begin{equation}
 K[v_{d,\tau}(x,y);x_1,x_2] = g(d,\tau)v_{d,\tau}(x_1,x_2) \label{fermionic-integral-equation}
\end{equation}
with eigenvalue $g(d,\tau)=1$. (Since the integration kernel is independent of $x_3$, we take the limit $|x_3|\rightarrow \infty$, which can also be obtained using a conformal transformation.) 


%
%
%
%
%

\section{Numerical Spectrum of Scalar Bilinears}
\label{numerical}

\subsection{Spectrum of Parity-Odd Scalar Bilinears}
Substituting the parity-odd scalar eigenvector ansatz \eqref{parity-odd-fermionic-eigenvectors} into the integral equation \eqref{fermionic-integral-equation}, gives the following expression for $g$ (see Appendix \ref{parity-odd-calculation} for details):
\begin{align}
 g_{\text{odd}}(d,\tau) = -\f{\G\big(\f{3d}4+\f12\big)\G\big(\f{d}4-\f{\tau}2\big)\G\big(\f{\tau}2-\f{d}4\big)}{\G\big(\f12-\f{d}4\big)\G\big(\f{d}4+\f{\tau}2\big)\G\big(\f{3d}4-\f{\tau}2\big)}. \label{g-odd-scalar}
\end{align}
Interestingly, this equation is independent of the ratio between $\lambda_1$ and $\lambda_2$.
Note that this equation \eqref{g-odd-scalar} reproduces equation 3.29 of \cite{Klebanov:2016xxf} when $d=1$.  

Solving the equation 
\be 
g_{\text{odd}}(d,\tau)=1,
\ee 
for $\tau$ will give us the scaling dimensions $\tau^{\text{(odd)}}_n$ of operators of the schematic form $\bar{\psi}\slashed{\partial}^{2n}\psi$. We expect $\tau^{\text{(odd)}}_n = 2n+2\Delta_\psi + \delta_n=2n+d/2+2\delta_n$ where $\delta_n \rightarrow 0$ as $n\rightarrow \infty$. 

\begin{figure}[h]
\begin{center}
\includegraphics[width=8cm]{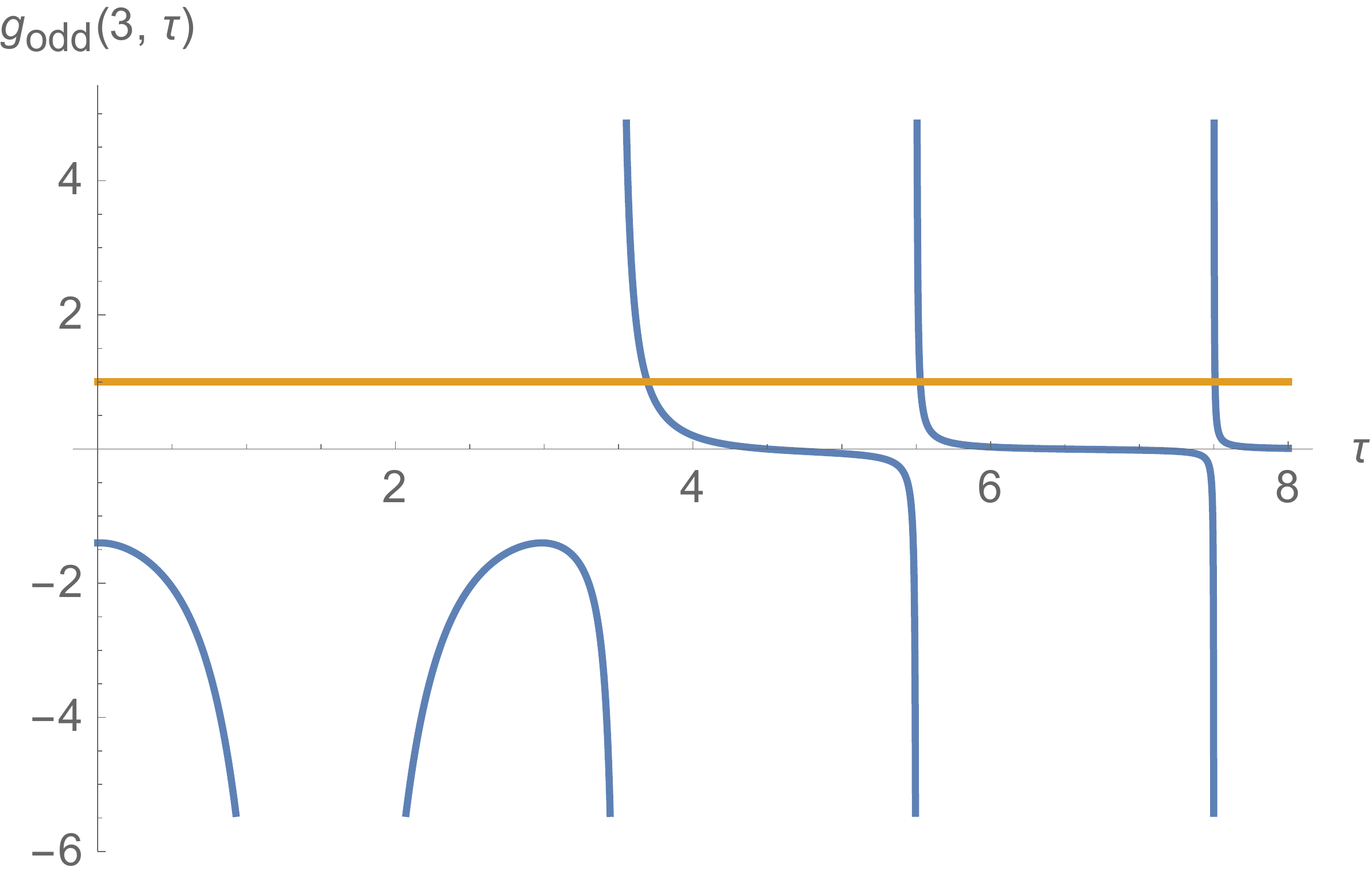}
\caption{A plot $g_{\text{odd}}(3,\tau)$ and $1$ for $d=3$. \label{odd-scalar-3}}
\end{center}
\end{figure}

Let us solve the equation $g_{\text{odd}}(d,\tau)=1$ for $\tau$, numerically when $d=3$. The solutions of this equation determine the allowed values of $\tau=\Delta$ for scalar operators in the large $N$ conformal fixed point we are studying. The plot of $g_{\text{odd}}(3,\tau)$ is shown in Figure \ref{odd-scalar-3}. 
The first few real roots that we find are: $\tau^{\text{(odd)}}_1=3.69364$, $\tau^{\text{(odd)}}_2=5.52725$, $\tau^{\text{(odd)}}_3=7.50793$, $\tau^{\text{(odd)}}_4=9.50331$. These approach $\tau^{\text{(odd)}}_n =2n+1.5$, as $n \rightarrow \infty$, as expected. A real eigenvalue corresponding to $n=0$ appears not to be present in the spectrum, but probably corresponds to the complex eigenvalue $\tau^{\text{(odd)}}_0  = 1.5 \pm 1.16817i$. The presence of this complex eigenvalue suggests the theory is unstable, and any putative gravitational dual description would contain fields below the BF bound, as discussed in \cite{Giombi:2017dtl}.

In $d=4$,  $g_{\text{odd}}(d,\tau)$ simplifies considerably:
\begin{equation}
g_{\text{odd}}(4,\tau)=\frac{15}{(\tau -4) (\tau -2)^2 \tau
   }.
\end{equation}
The roots to $g_{\text{odd}}(4,\tau)=1$ are: 
\be 
\tau^\text{(odd)}=\left\{2-i \sqrt{\sqrt{19}-2},~ 2+i
   \sqrt{\sqrt{19}-2}, ~ 2-\sqrt{2+\sqrt{19}}, ~ 2+\sqrt{2+\sqrt{19}}\right
   \}.
\ee
Interestingly, there is no tower of solutions for this case (which resembles the $d=2$ case in the bosonic tensor model of \cite{Giombi:2017dtl}.) We again find complex solutions indicating the theory is unstable. We also find a parity-odd complex eigenvalue in all dimensions $2<d<6$. 

\begin{figure}[h]
\begin{center}
\includegraphics[width=8cm]{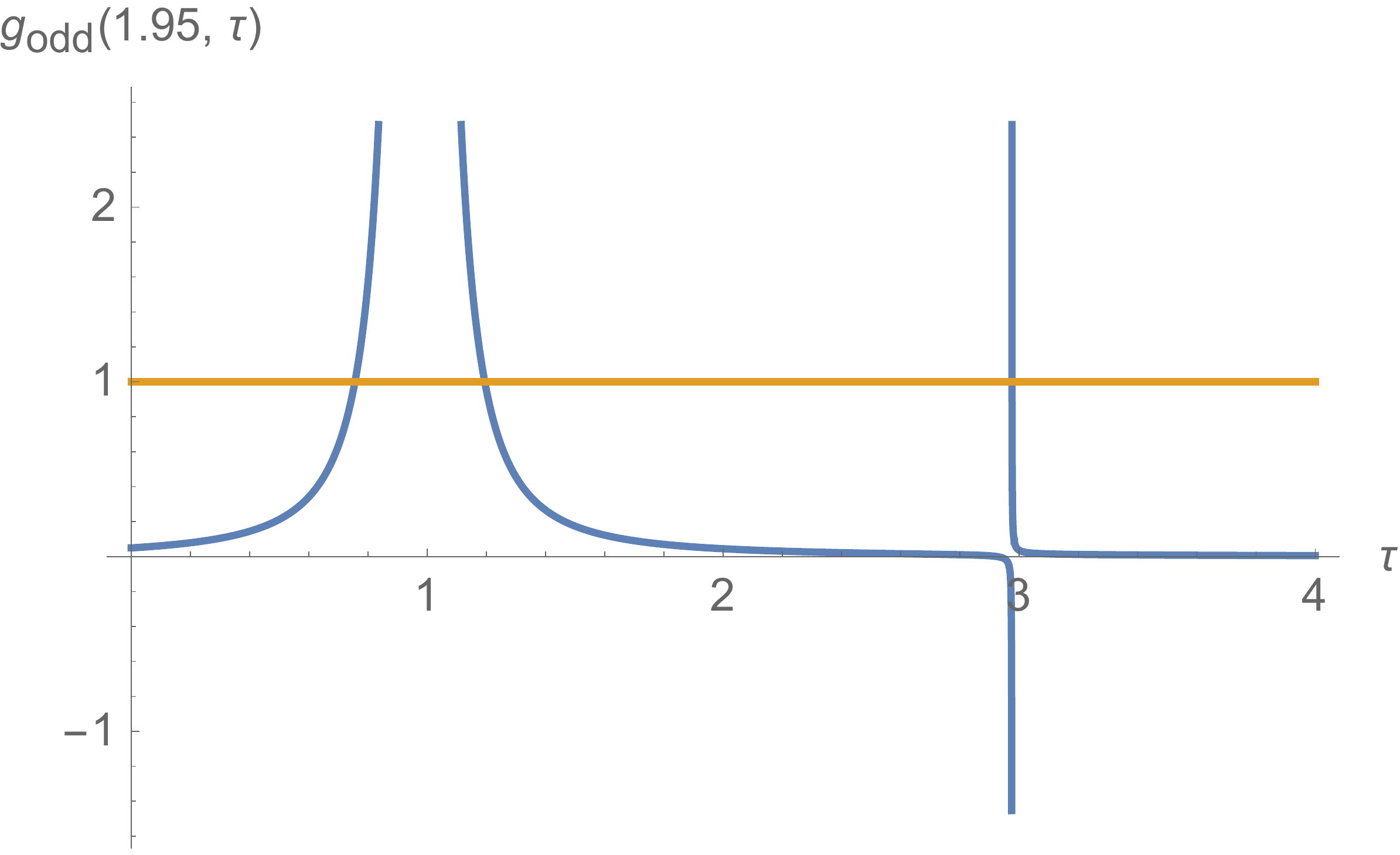}
\includegraphics[width=8cm]{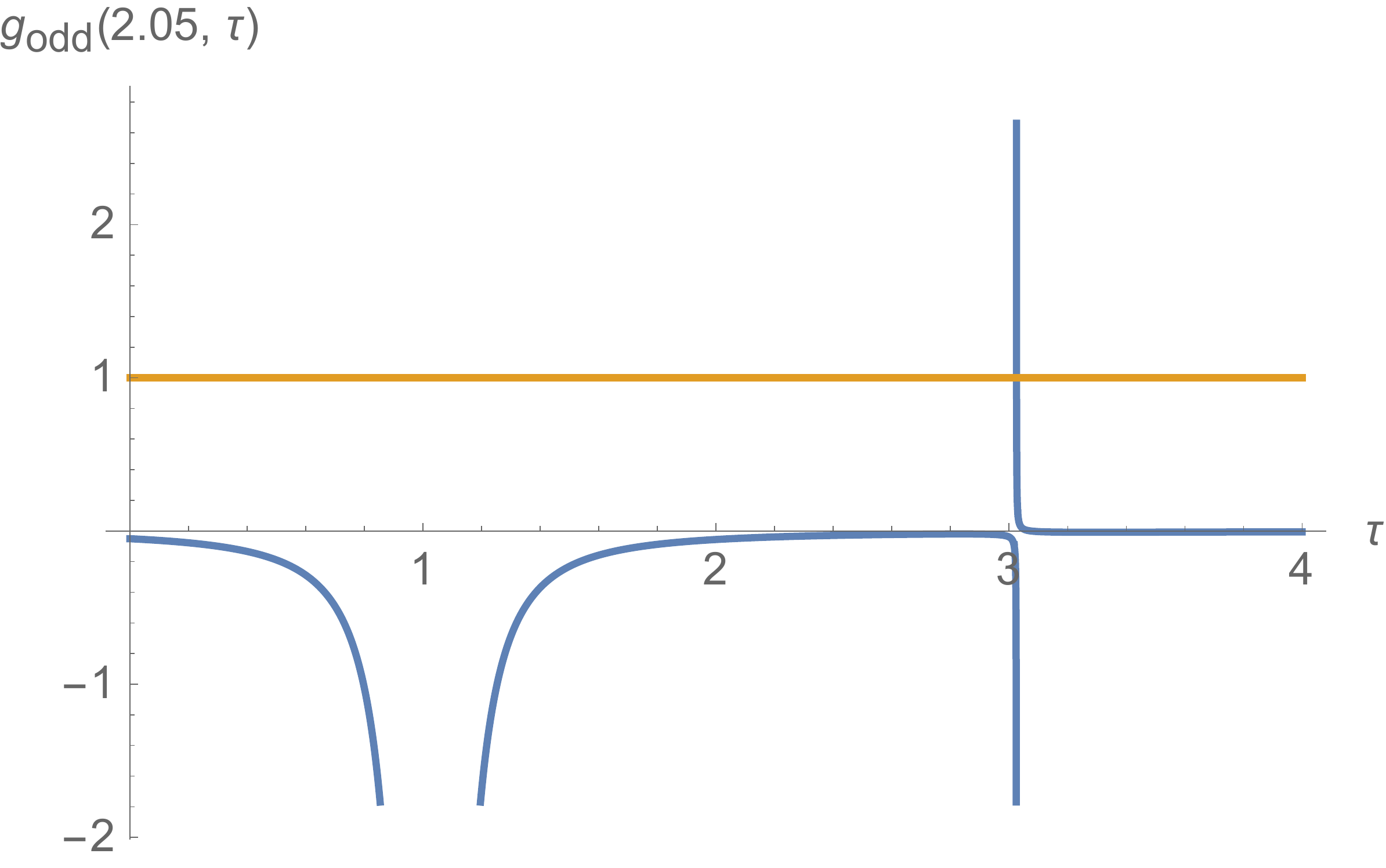}
\caption{A plot $g_{\text{odd}}(\tau)$ and $1$ for $d=1.95$ (top) and $d=2.05$ (below). \label{2epsilon}}
\end{center}
\end{figure}

In $d<2$ dimensions, based on a numerical search we conclude that the spectrum does not contain any complex eigenvalue, while in $d>2$ (but $d<6$) dimensions the spectrum does contain a complex eigenvalue. A plot in $d=1.95$ and $d=2.05$ is shown in the figure \ref{2epsilon}, below, which illustrates this fact. We also note that $g_{\text{odd}}(2,\tau)$ vanishes, indicating that the conformal fixed point is free in $2$ dimensions.  The first few eigenvalues in $d=1.95$ are $0.755708$, $1.19429$, $2.97562$, $4.97516$, $6.97507$. The first few eigenvalues in $d=2.05$ are $1.025 \pm 0.227675 i$, $3.02562$, $5.02515$, $7.02507$. We present analytic expressions for these eigenvalues as a power series in $\epsilon$ in $d=2-\epsilon$ dimensions in the section \ref{2-epsilon}.  

\begin{figure}[h]
\begin{center}
\includegraphics[width=8cm]{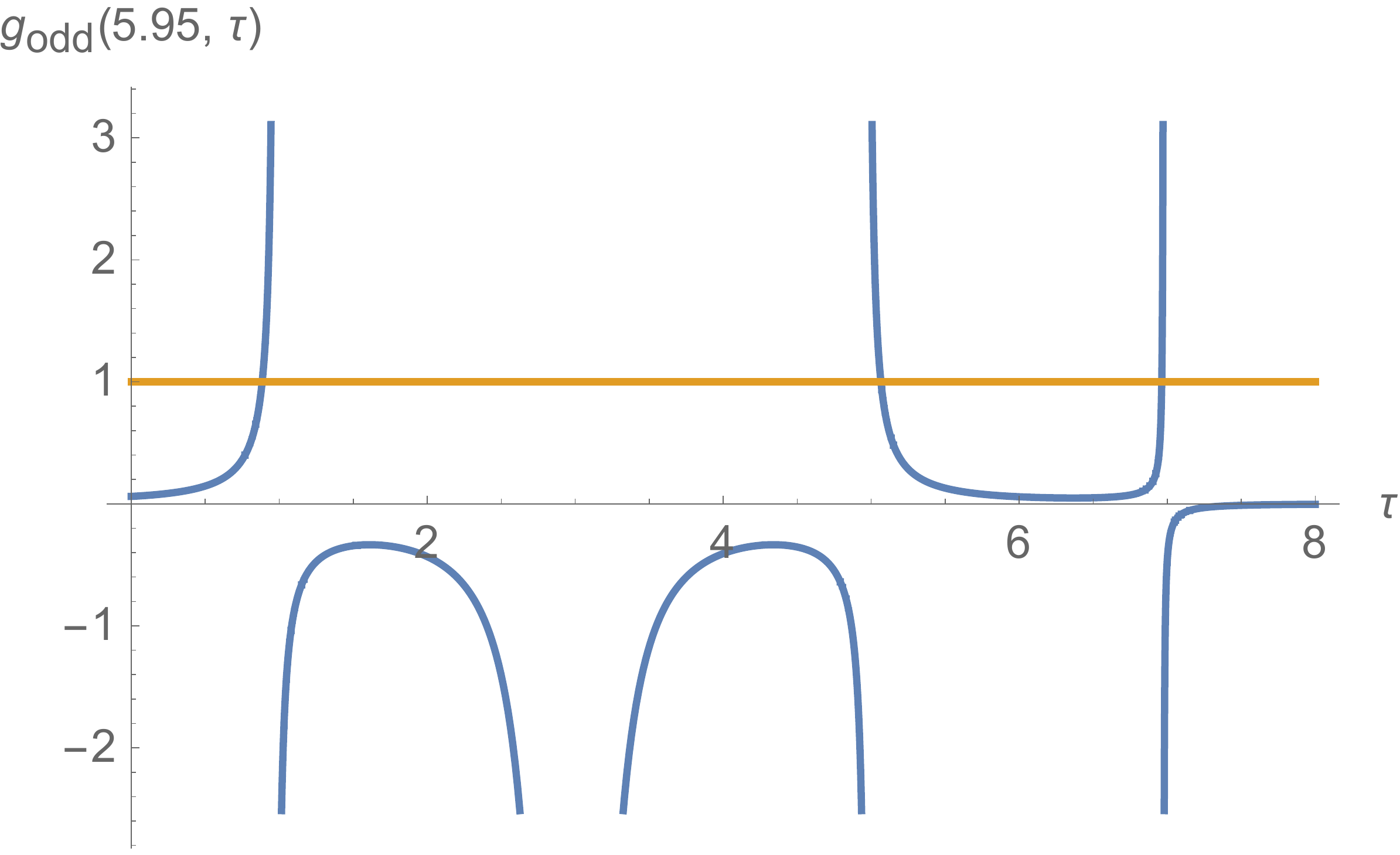}
\includegraphics[width=8cm]{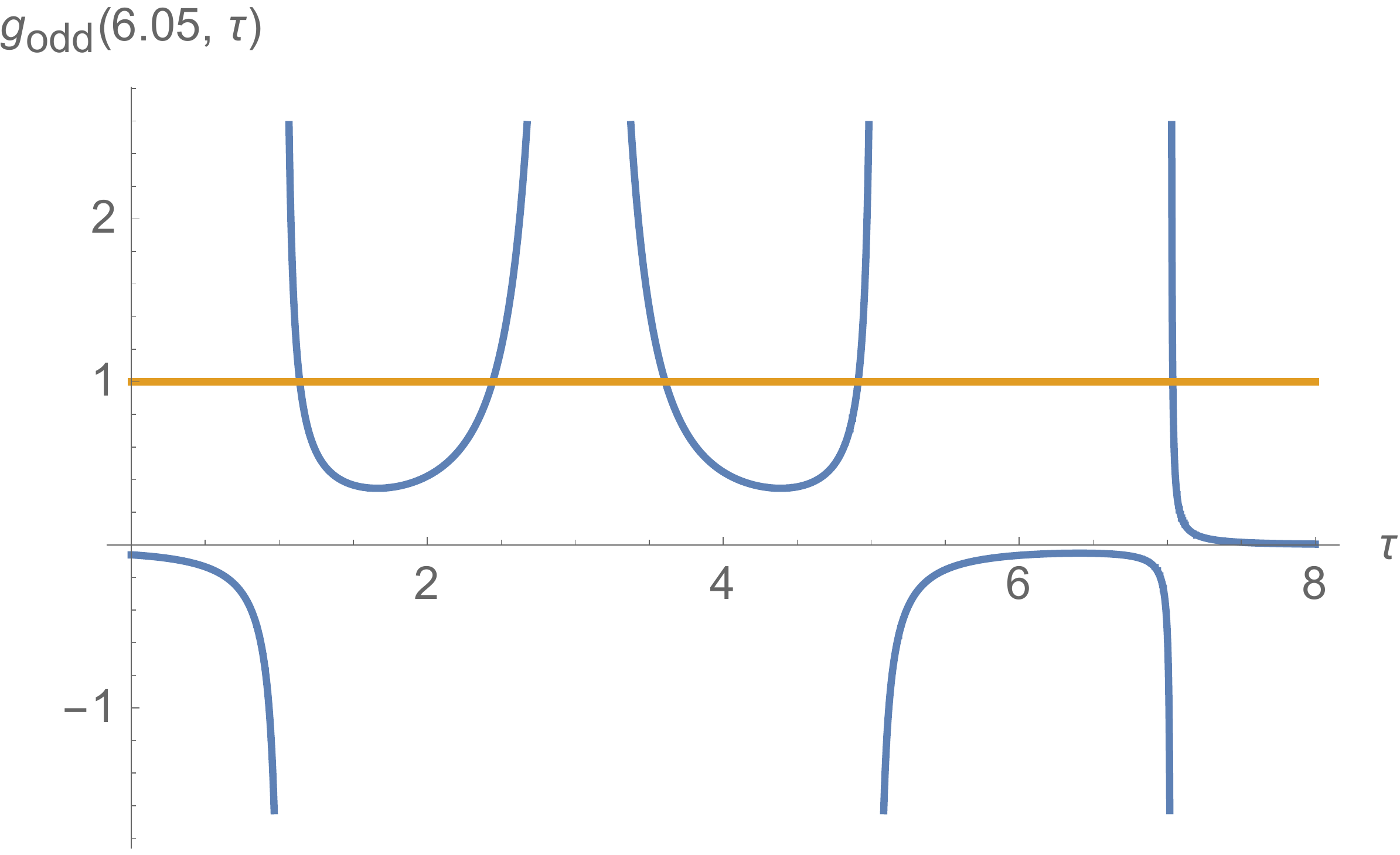}
\caption{A plot $g_{\text{odd}}(\tau)$ and $1$ for $d=5.95$ (top) and $d=6.05$ (below). We see that some eigenvalues become real as one crosses $d=6$. \label{6epsilon}}
\end{center}
\end{figure}

Note that $g_{\text{odd}}$ also vanishes for $d=6$ (and $d=4n+2$) in general, suggesting the (UV) conformal fixed point is free in these dimensions as well. Because $g(d,\tau)$ changes sign at $d=6$ (as shown in the figure \ref{6epsilon}, below) the spectrum is qualitatively different in $6-\epsilon$ dimensions, which contains a complex eigenvalue, and in $6+\epsilon$ dimensions, which seems to have a purely real spectrum. We find that $\tau=2.995 - 0.242346 i$ is a complex eigenvalue for $d=5.99$.

\begin{figure}[h]
\begin{center}
\includegraphics[width=8cm]{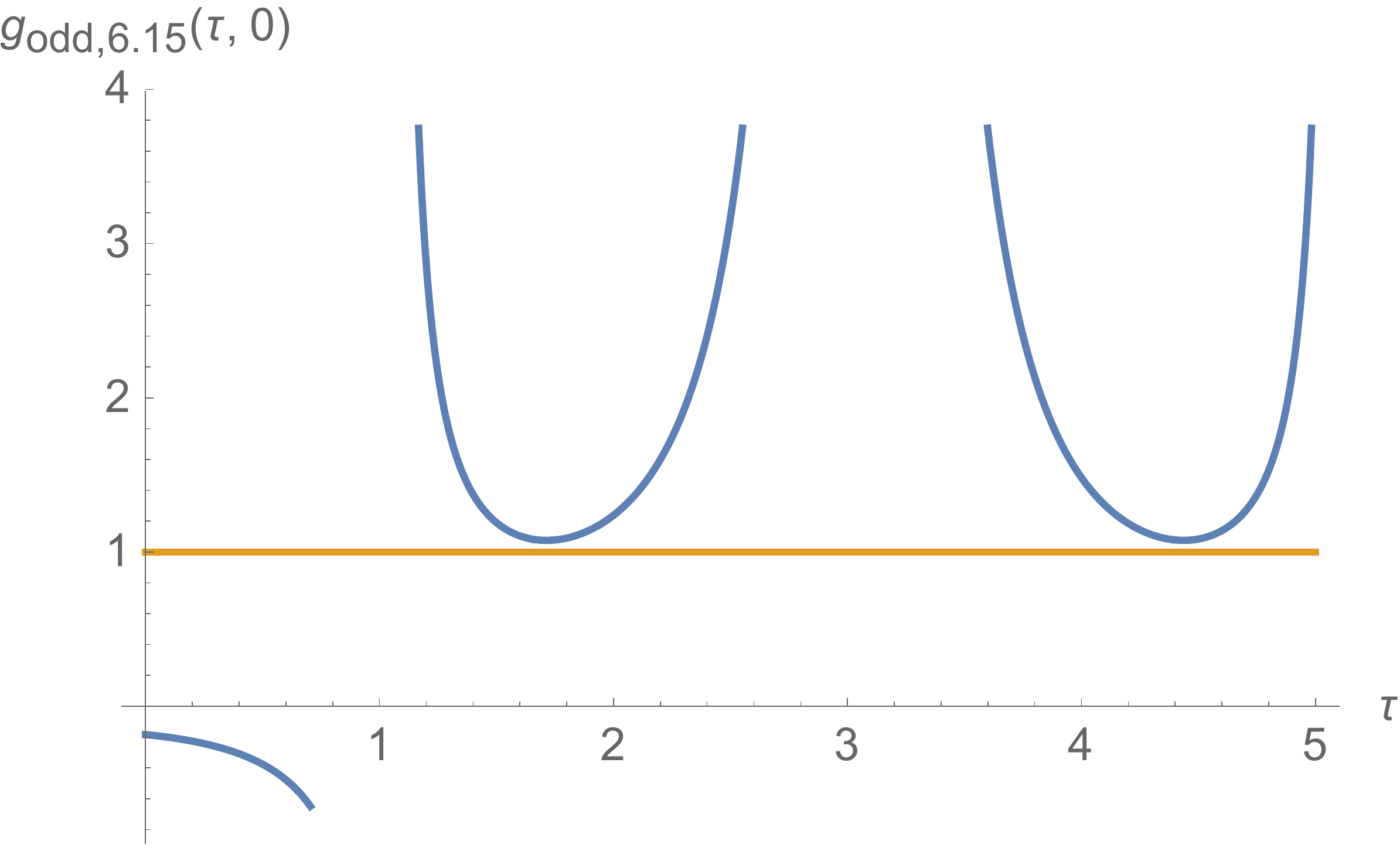}
\includegraphics[width=8cm]{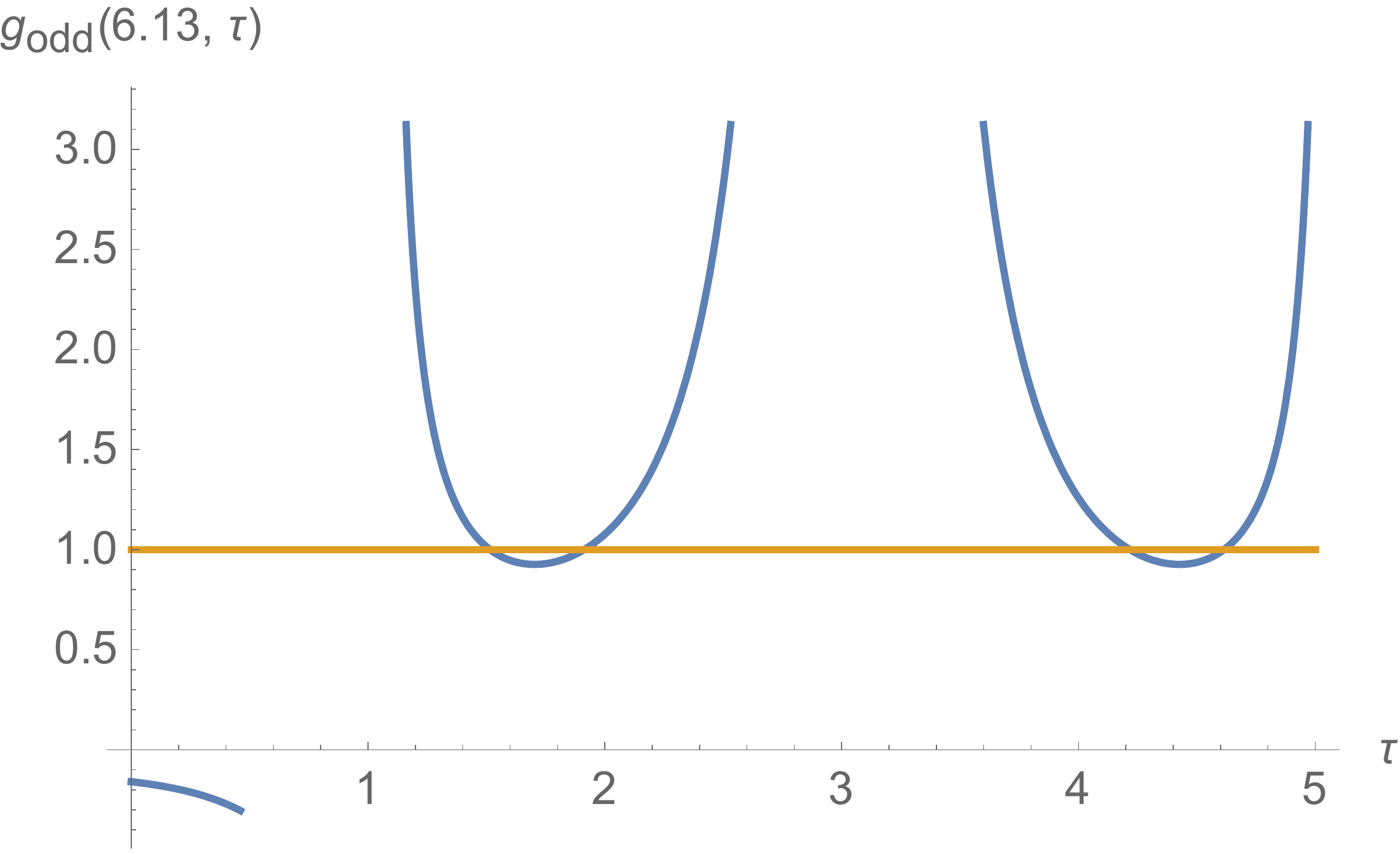}
\caption{A plot $g_{\text{odd}}(\tau)$ and $1$ for $d=6.15$ (top) and $d=6.13$ (below) indicating that some eigenvalues become real below $d<6.14$. \label{6epsilon2}}
\end{center}
\end{figure}

Studying the case of $d=6+\epsilon$ dimensions numerically, we find $\tau=1.70838 - 0.0178181 i$ is a complex eigenvalue for $d=6.14$. This complex eigenvalue persists for higher values of $d$ but disappears when $d<6.13$, as shown figure \ref{6epsilon2}, below. Numerically, we cannot find a complex eigenvalue for $6.14>d>6$, so it may be possible to define an interacting melonic theory free from complex eigenvalues in this range of dimensions. For example, the first few numerical eigenvalues in $6.05$ dimensions are $1.13874$, $2.44164$, $3.60836$, $4.91126$, $7.03777$, $9.02496$, $11.025$. In section \ref{6+epsilon} below, we analytically compute the spectrum in $6+\epsilon$ dimensions as a power series in $\epsilon$, and verify it is real when $\epsilon$ is positive. Unlike the case $d<2$, the first few eigenvalues listed above in the spectrum appear to be below the unitary bound for scalars in 6 dimensions ($\tau_*=2$). If the theory is unitary in $d=6+\epsilon$ then these eigenvalues must be spurious, and the spectrum begins at $\tau=2.44$, which naturally corresponds to the operator $\bar{\psi}\slashed{\partial}^{2n}\psi$, with $n=0$. 

While we expect complex eigenvalues for generic values of $d>6.14$, there may be additional ``windows"  at larger values of $d$ where the spectrum is real, similar to the range $6<d<6.14$.

\subsection{Spectrum of Parity-Even Scalar Bilinears}
Substituting in the  parity-even eigenvector \eqref{parity-even-fermionic-eigenvectors}, into the integral equation \eqref{fermionic-integral-equation}, gives the following (see Appendix \ref{parity-even-calculation} for details):
\begin{equation}
\begin{split}
g_{\text{even}}(d,\tau) & = -\frac{3 \cos \left(\frac{\pi  d}{4}\right) \Gamma \left(\frac{3 d}{4}+\frac{1}{2}\right) \Gamma \left(\frac{d+2}{4}\right) \sec \left(\frac{1}{4} \pi 
   (d-2 \tau )\right)}{\Gamma \left(\frac{1}{4} (3 d-2 \tau +2)\right) \Gamma \left(\frac{1}{4} (d+2 \tau +2)\right)}
   \end{split}
\end{equation}
As in the previous section, this expression is independent of the ratio between $\lambda_1$ and $\lambda_2$, and we must solve 
\be
g_{\text{even}}(d,\tau)=1
\ee
to determine the scaling dimensions of operators of the schematic form $\bar{\psi}\slashed{\partial}^{2n+1}\psi$. We expect eigenvalues of the form $\tau^{\text{(even)}}_n=(2n+1)+2 \Delta_\psi+\delta_n=2n+1+\frac{d}{2} +\delta_n$, with $\delta_n \rightarrow 0$ as $n \rightarrow \infty$. 

\begin{figure}[h]
\begin{center}
\includegraphics[width=8cm]{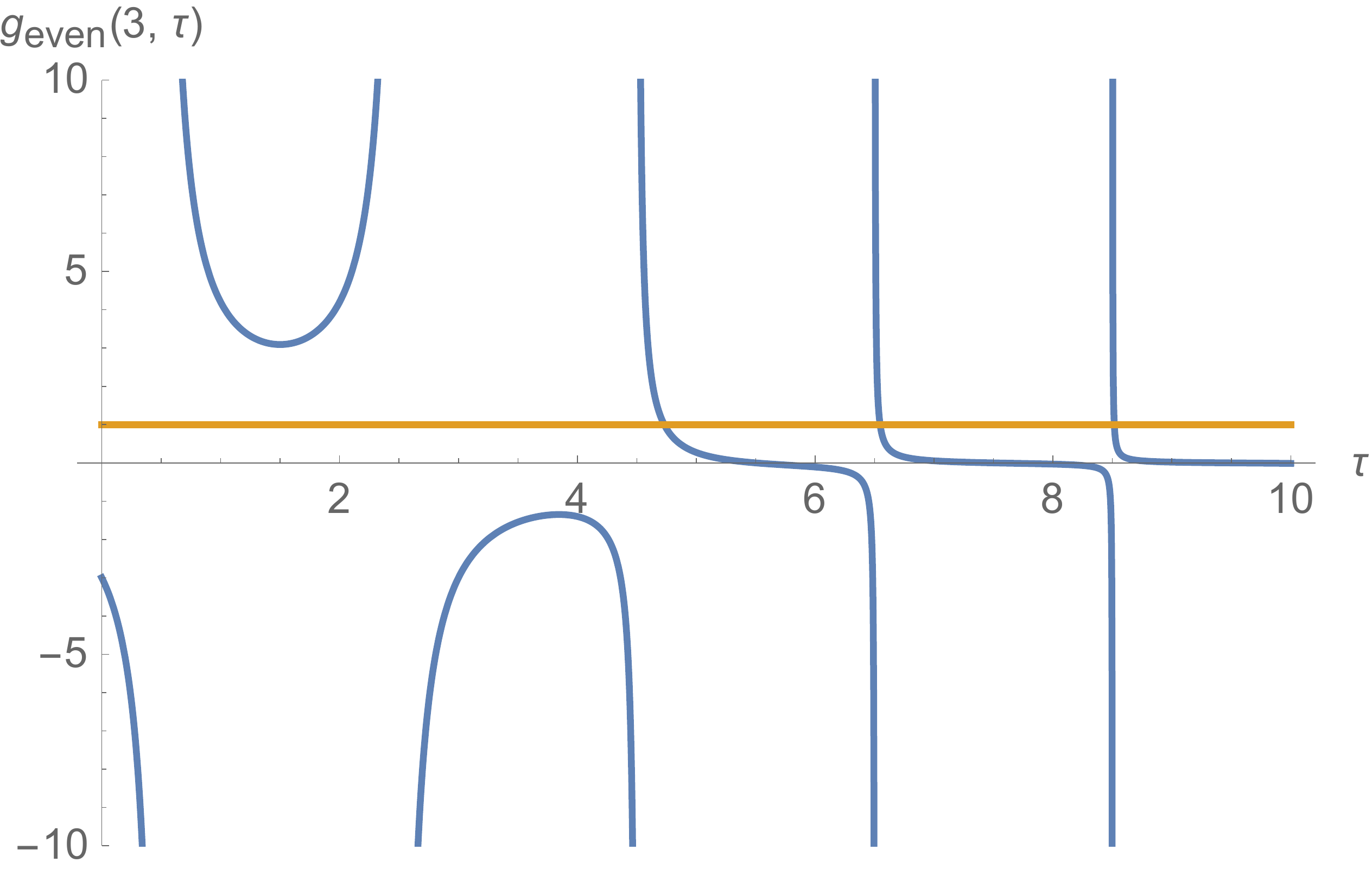}
\caption{A plot $g_{\text{even}}(\tau)$ for $d=3$. \label{even-scalar-3}}
\end{center}
\end{figure}

For $d=3$, the plot of $g$ is shown in Figure \ref{even-scalar-3}, the lowest solutions to $g_{\text{even}}(\tau,0)=1$ are $\tau^{\text{(even)}}_1=4.73049$, $\tau^{\text{(even)}}_2=6.5462$, $\tau^{\text{(even)}}_3=8.5158$, $\tau^{\text{(even)}}_4=10.5072$, $\tau^{\text{(even)}}_5=12.5039$. These approach $2n+2.5$ as expected, though $n=0$ is missing. There is a complex solution $1.5 - 1.32587i$, which likely corresponds to $n=0$. 

\begin{figure}[h]
\begin{center}
\includegraphics[width=8cm]{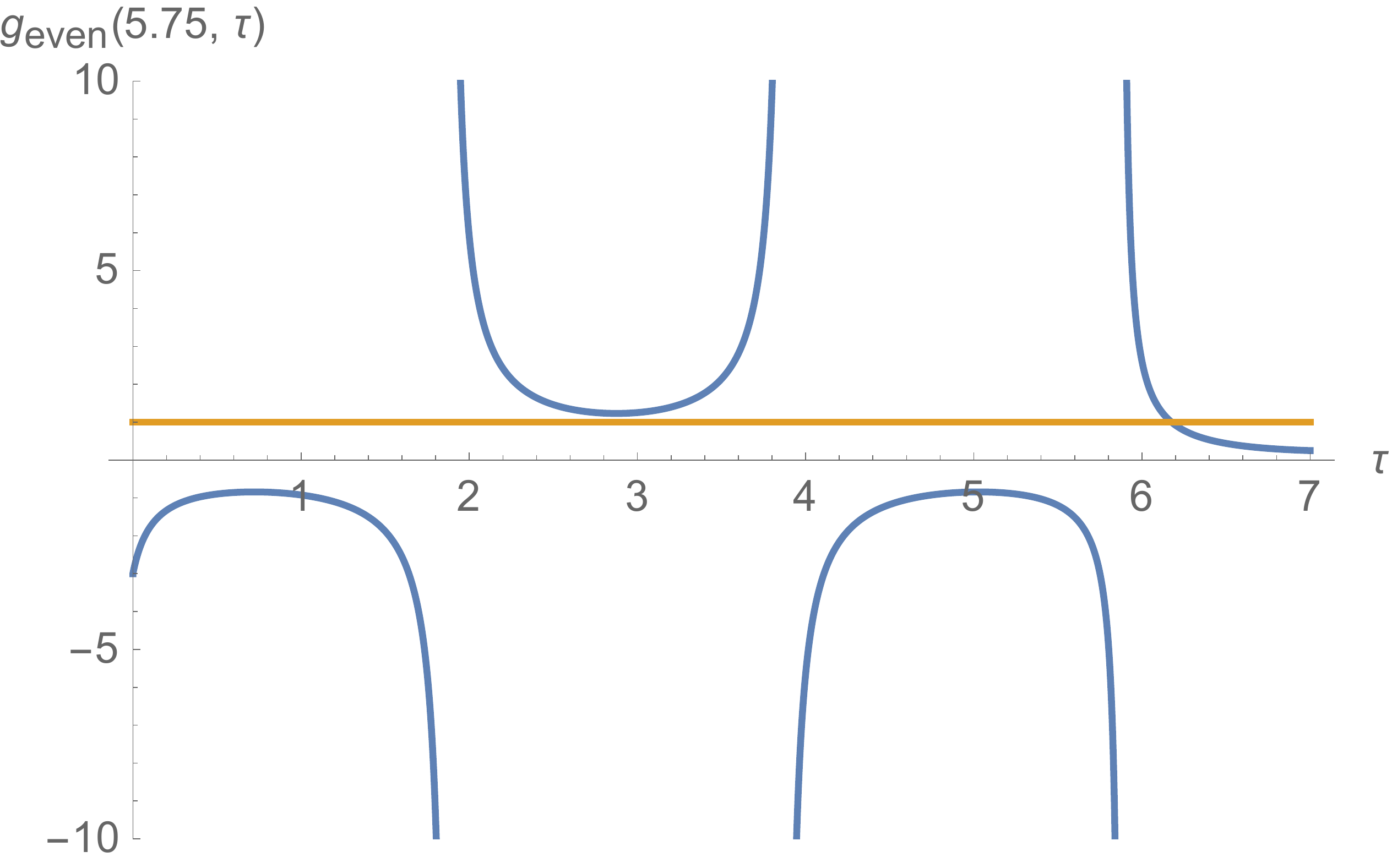}
\includegraphics[width=8cm]{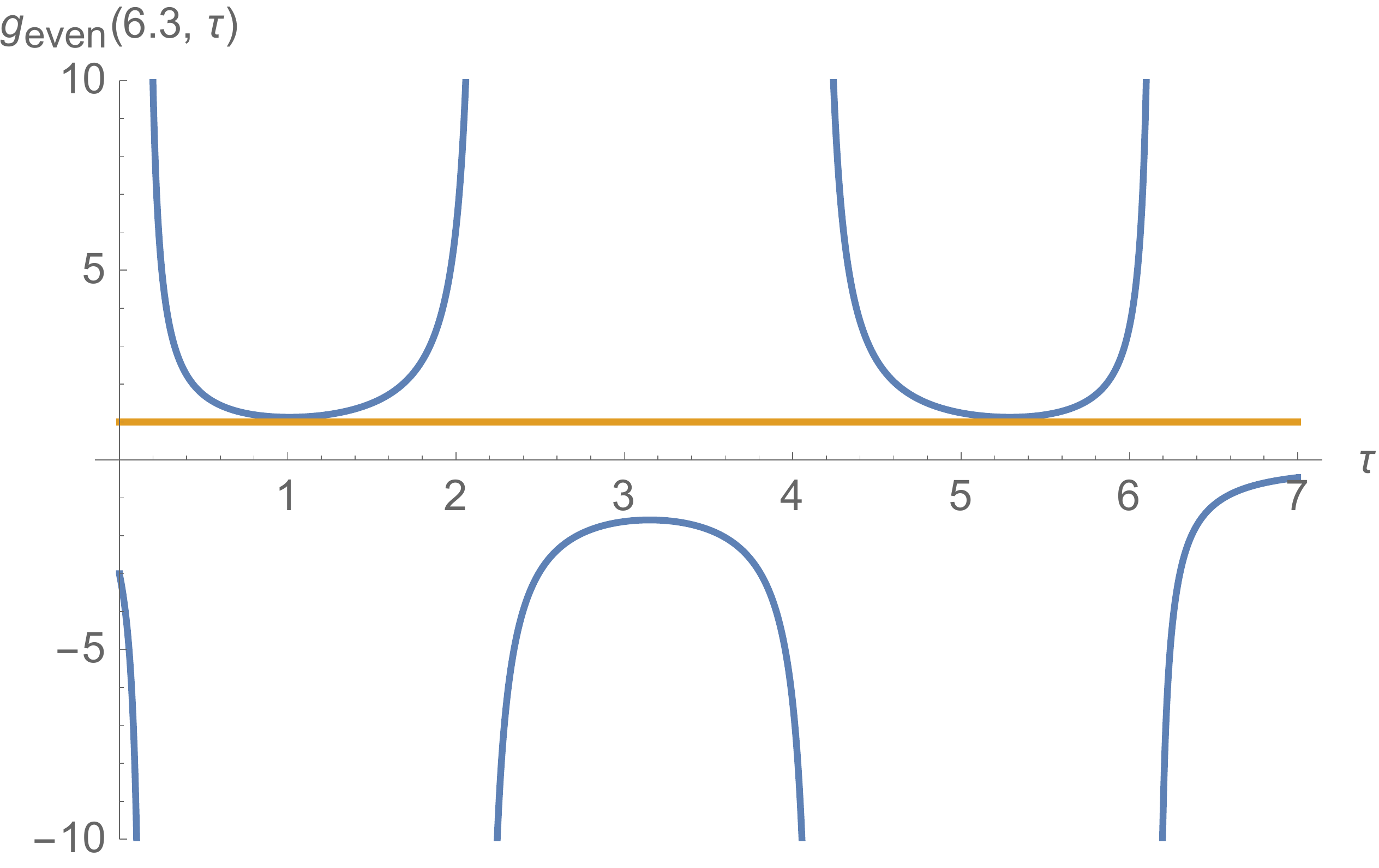}
\caption{A plot $g_{\text{even}}(\tau)$ and $1$ for $d=5.75$ (top) and $d=6.30$ (below) indicating that some eigenvalues become complex outside the range $5.79<d<6.26$. \label{575}}
\end{center}
\end{figure}

Performing a numerical search for complex eigenvalues in dimensions less than $7$, we only find a parity-even complex eigenvalue in the range $2.3225<d<5.79$, and $d>6.26$ -- which is a subset of the range for which their exists a parity-odd complex eigenvalue. (As pictured in figure \ref{575}, for $d=5.75$, $\tau=2.875 + 0.442 i$ is a complex eigenvalue, and for $d=6.30$, $\tau=0.998 + 0.317 i$ is a complex eigenvalue.) These results is consistent with the conjecture that the theory contains only real eigenvalues in $d=6+\epsilon$. 

When $d=4$, $g_{\text{even}}(d,\tau)$ takes a simple form:
\begin{equation}
g_{\text{even}}(4,\tau) = \frac{45}{\tau ^4-8 \tau ^3+14 \tau ^2+8 \tau
   -15}
\end{equation}
which corresponds to the eigenvalues: 
\be
\tau^{\text{(even)}}=\left\{2-i \sqrt{\sqrt{61}-5},2+i
   \sqrt{\sqrt{61}-5},2-\sqrt{5+\sqrt{61}},2+\sqrt{5+\sqrt{61}}\right\}.
\ee 
Again, there is no tower of solutions in this case.

We also note that $g_{\text{even}}(d,\tau)$ vanishes when $d=2$ and $d=6$, as expected from the analysis of $g_{\text{odd}}$, consistent with the claim that the fixed point is free in these dimensions.

\begin{figure}[h]
\begin{center}
\includegraphics[width=8cm]{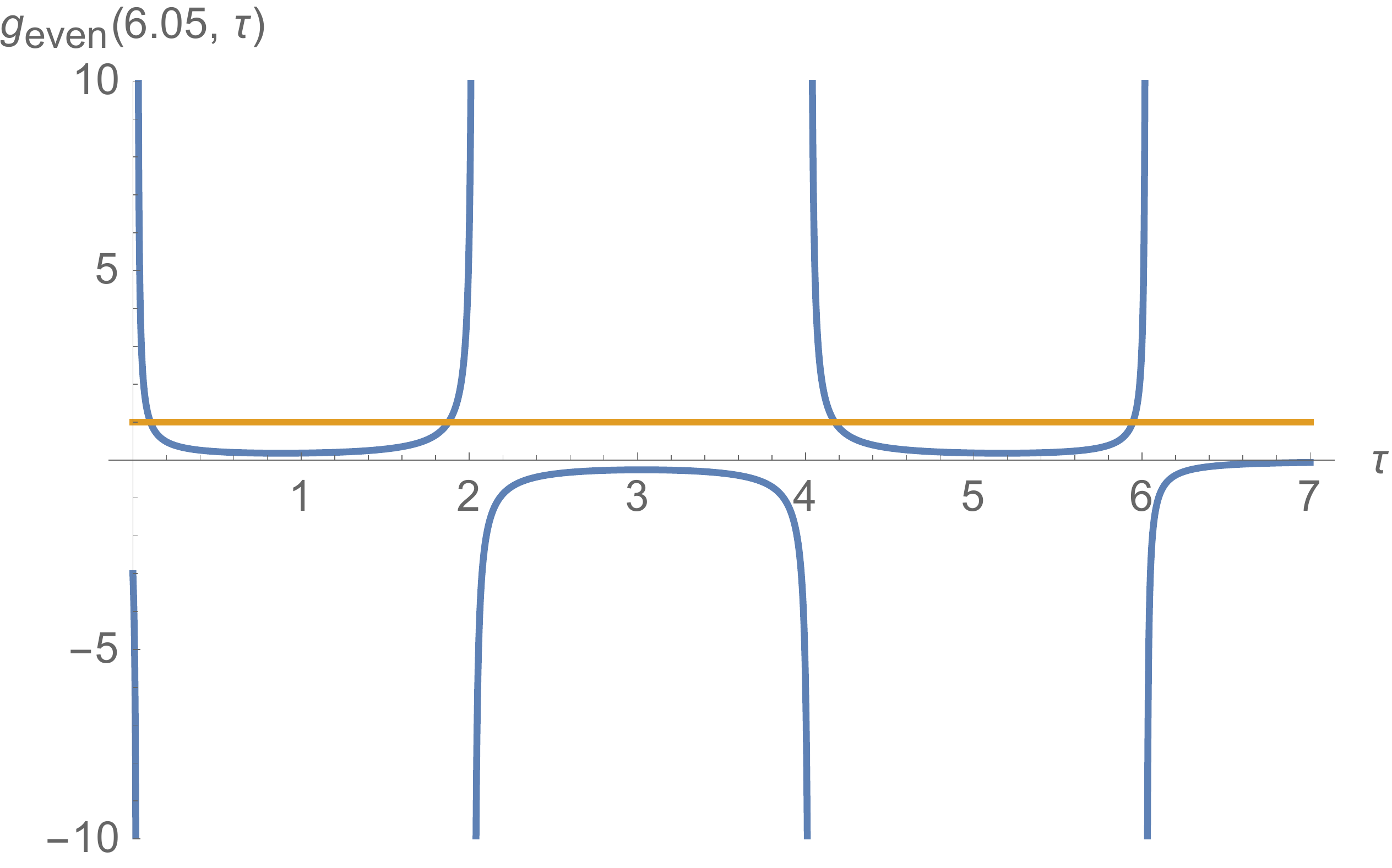}
\caption{A plot $g_{\text{even}}(\tau)$ for $d=6.05$. \label{even605}}
\end{center}
\end{figure}

For $d=6.05$, plotted in Figure \ref{even605}, the first few eigenvalues are $0.10446$, $1.87654$, $4.17346$, $5.94554$, $8.04024$,     $10.0249$,$\ldots$. The first two of these eigenvalues lie below the unitary bound. If the theory is unitary in $d=6+\epsilon$ then these eigenvalues must be spurious, and the spectrum begins at $\tau=4.17$, which naturally corresponds to the operator $\bar{\psi}\slashed{\partial}^{2n+1}\psi$, with $n=0$.

\section{Epsilon Expansion in $d=2-\epsilon$ and $d=6+\epsilon$}

Numerically, we found that the spectrum of scalar bilinears is real when $d<2$ and when $6.14>d>6$, and the scaling dimensions approach free values as $d\rightarrow 2$ or $6$. In this section, we will present analytic expressions for these real scaling dimensions in $2-\epsilon$ and $6+\epsilon$ dimensions, obtained by solving the equations $g(2-\epsilon,\tau)=1$ and $g(6+\epsilon,\tau)=1$ perturbatively in $\epsilon$. Of course, the case of $d=6$ is not necessarily physical. 

In the expressions that follow, $H_n$ denotes the $n$th harmonic number, 
$H_n=\displaystyle \sum_{k=1}^n \frac{1}{k}.$

\subsection{$d=2-\epsilon$}

\label{2-epsilon}
When $d=2-\epsilon$, we appear to have a sensible IR fixed point, with a purely real spectrum in the large $N$ limit. All operators appear to have scaling dimensions above the unitary bound. 

Solving $g_{\text{odd}}(2-\epsilon,\tau)=1$, we find the parity-odd scalar spectrum in $d=2-\epsilon$ is:
\begin{eqnarray}
\tau^{\text{(odd)}}_{0,\pm} & = & 1 \pm \sqrt{\epsilon}-\frac{1}{2}\epsilon \pm \frac{3}{8}\epsilon^{3/2} 
					\pm \left(\frac{\zeta(3)}{8}+\frac{9}{128}\right) \epsilon^{5/2}
					+ O\left(\epsilon^3\right) 
					\\
\tau^{\text{(odd)}}_1 & = & 3 - \frac{1}{2}\epsilon 
				+\frac{1}{4}\epsilon^2 + 0 \epsilon^3
				-\frac{7}{64}\epsilon^4+O\left(\epsilon^5\right) \\
\tau^{\text{(odd)}}_n & = & (2 n+1)-\frac{1}{2}\epsilon
	+\frac{1}{4n^2}\epsilon^2
   	+\frac{\left(4 n^2 H_{n-1}+(2-3 n)n+1\right)}{16n^4}\epsilon^3 \\
   	& & 
   	+\frac{\left(8 n^4 H_{n-1}^2-6n^3-2 n^2+4 ((2-3 n) n+2)n^2 H_{n-1}+1\right)}{64n^6}\epsilon ^4
   	+O\left(\epsilon^5\right) \text{ for $n \geq 1$.} \nonumber
\end{eqnarray}

The parity-even scalar spectrum is:
\begin{eqnarray}
\tau^{\text{(even)}}_0 & = & 2+\epsilon -\frac{3 \epsilon ^2}{2}+\frac{3
   \epsilon ^3}{2}+\left(-\frac{3 \zeta
   (3)}{4}-\frac{21}{8}\right) \epsilon
   ^4+O\left(\epsilon ^5\right) \\
\tau^{\text{(even)}}_n & = & 2 (n+1)
             -\frac{1}{2}\epsilon 
             +\frac{3 }{4 n (n+1)}\epsilon ^2
             +\frac{3  \left(4(n+1) n H_{n-1}-3 n^2+n+5\right)}{16 n^2 (n+1)^2} \epsilon ^3 \\ &&
             +\frac{3  \left(8 (n+1)^2n^2 H_{n-1}^2
             +4 (n (9-n (3 n+2))+8) n H_{n-1}-12 n^3-24n^2+19\right)}{64 n^3 (n+1)^3}\epsilon ^4
             +O\left(\epsilon ^5\right), \text{ for $n\geq1$.} \nonumber
\end{eqnarray}

   
\subsection{$d=6+\epsilon$}
\label{6+epsilon}
We find a purely real spectrum in $d=6+\epsilon$ dimensions for $\epsilon<0.14$. This spectrum may contain eigenvalues below the unitary bound, and the colored fermions $\psi^{abc}$ in this dimension have dimension $3/2$ which is also below the unitary bound $5/2$. As such, the formal large $N$ fixed point in this case may be non-unitary or otherwise ill-defined. Nevertheless, it is remarkable that there is a small window around $d=6$ in which the spectrum is real, so we present some results below.

The parity-odd scalar spectrum is:
\begin{eqnarray}
\tau^{\text{(odd)}}_{-1} & = & 1+\frac{5}{2}\epsilon  +\frac{107}{24}\epsilon^2 + 
				\frac{3047}{192} \epsilon^3 + 
				\left(\frac{15 \zeta (3)}{8}+\frac{484679}{6912}\right) \epsilon^4+O\left(\epsilon ^5\right)\\
\tau^{\text{(odd)}}_{0,\pm} & = & 3 \pm \sqrt{6} \sqrt{\epsilon} + \frac{1}{2}\epsilon
					 \pm \frac{35}{16}\sqrt{\frac{3}{2}} \epsilon^{3/2}
					\pm \frac{\left(1536 \sqrt{6}\zeta (3)+4799 \sqrt{6}\right)}{2048}\epsilon^{5/2}
					+O\left(\epsilon^{7/2}\right) \\
\tau^{\text{(odd)}}_{1} & = & 5 - \frac{3}{2}\epsilon -\frac{107}{24}\epsilon^2-\frac{3047}{192}\epsilon^3
			  +\left(-\frac{15 \zeta(3)}{8}-\frac{484679}{6912}\right) \epsilon ^4+O\left(\epsilon^5\right) 
			  \\
\tau^{\text{(odd)}}_{2} & = & 7+\frac{3}{4}\epsilon +\frac{43}{384}\epsilon^2-\frac{301}{6144}\epsilon^3
				+\left(\frac{63713}{3538944}-\frac{3 \zeta (3)}{256}\right) \epsilon^4
				+O\left(\epsilon ^5\right)\\
\tau^{\text{(odd)}}_{2+n} & = & (2 n+7)+
				\frac{1}{2}\epsilon
				-\frac{6\Gamma (n)}{(n+2) \Gamma (n+5)}\epsilon^2
 				\\ && 
 				+\frac{ 6\left(\frac{1}{2}\left(\frac{1}{n+1}+\frac{1}{n+2}+\frac{1}{n+3}+\frac{1}{n+4}+\frac{1}{n}\right)
 						+H_{n-1}
 						-\frac{6 \Gamma (n) \Gamma(n+2)}{\Gamma (n+3) \Gamma (n+5)}
 						-\frac{29}{16}\right)}
 				{n (n+1)(n+2)^2 (n+3) (n+4)}
 				\epsilon ^3
 				+O\left(\epsilon^4\right) \text{, for $n\geq1$.} \nonumber
\end{eqnarray}

The parity-even scalar spectrum is
\begin{eqnarray}
\tau^{\text{(even)}}_{-2} & = & 2 \epsilon +\frac{13 \epsilon ^2}{8}+\frac{67 \epsilon ^3}{24}+\left(\frac{3 \zeta
   (3)}{4}+\frac{54401}{9216}\right) \epsilon ^4+O\left(\epsilon ^5\right) \\
\tau^{\text{(even)}}_{-1} & = & 2-\frac{5 \epsilon }{2}+\frac{17 \epsilon ^2}{16}-\frac{3635 \epsilon
   ^3}{384}+\left(\frac{50759}{2304}-\frac{105 \zeta (3)}{16}\right) \epsilon
   ^4+O\left(\epsilon ^5\right) \\
\tau^{\text{(even)}}_0 &=&   4+\frac{7 \epsilon }{2}-\frac{17 \epsilon ^2}{16}+\frac{3635 \epsilon
   ^3}{384}+\left(\frac{105 \zeta (3)}{16}-\frac{50759}{2304}\right) \epsilon
   ^4+O\left(\epsilon ^5\right)
   \\
   \tau^{\text{(even)}}_1 & = & 6-\epsilon -\frac{13 \epsilon ^2}{8}-\frac{67 \epsilon ^3}{24}+\left(-\frac{3 \zeta
   (3)}{4}-\frac{54401}{9216}\right) \epsilon ^4+O\left(\epsilon ^5\right) 
   \\
   \tau^{\text{(even)}}_2 & = & 8+\frac{4 \epsilon }{5}+\frac{197 \epsilon ^2}{2000}-\frac{32581 \epsilon
   ^3}{600000}+\left(\frac{8429}{281250}-\frac{3 \zeta (3)}{250}\right) \epsilon
   ^4+O\left(\epsilon ^5\right) 
 \\
   \tau^{\text{(even)}}_{2+n} & = & 8+2n+\frac{\epsilon
   }{2}-\frac{18 \epsilon ^2
   \Gamma (n)}{\Gamma
   (n+6)}+ \nonumber \\ && \frac{9 \epsilon ^3
   \Gamma (n) \left(8 H_{n+5}
   \Gamma (n+6)-288 \Gamma
   (n)+\Gamma (n+6) (8 H_{n-1}-29)\right)}{8 \Gamma
   (n+6)^2}+O\left(\epsilon
   ^4\right) 
\end{eqnarray}


While this six-dimensional fixed point might not be physical, let us make a few brief comments about it.  

We labeled the eigenvalues above as $\tau^{\text{(even)}}_n$ if the scaling dimension at $\epsilon=0$ was equal to $2\Delta_\psi+(2n+1)$, corresponding to the operator $\bar{\psi}\slashed{\partial}^(2n+1)\psi$, and $\tau^{\text{(odd)}}_n$ if the scaling dimension at $\epsilon=0$ was equal to $2\Delta_\psi+(2n)$, corresponding to the operator $\bar{\psi}\slashed{\partial}^(2n)\psi$. We find some eigenvalues corresponding to negative values of $n$, listed above, and these are presumably not physical. If these eigenvalues are scaling dimensions of genuine bilinear operators, then even a gauged-version of the 6-dimensional theory (containing only singlets formed out of $\psi$ and $\bar{\psi}$ as gauge-invariant operators) would be non-unitary, since these operators have scaling dimensions below the unitary bound for (non-singleton) scalars in $d=6$, which is $\tau_*=(d-2)/2=2$. 

If these eigenvalues can be excluded, we still expect that the theory restricted to the singlet sector is not unitary, but to see this, one would have to look at the higher-spin spectrum. The theory in $6$ dimensions appears to be a theory of free fermions with non-standard scaling dimension $\frac{3}{2}$. The unitarity bound for a vector in $6$ dimensions is $5$ and $\bar{\psi}\gamma_\mu\psi$ would likely have scaling dimension $3$, which is well below the bound. These arguments apply to all the theories with $d>2$, but in dimensions such as $3$ where interactions are non-trivial, it might be possible that scaling dimensions of higher-spin currents could be lifted above the unitarity bound. While some constructions of formal theories containing negative mass higher-spin gauge fields in AdS do exist \cite{Brust:2016zns, Brust:2016xif}, it does appear that the existence of a gravitational dual for the formal UV fixed point in $d=6$ is unlikely.

\section*{Acknowledgements}

The authors thank Igor Klebanov for discussions and for reading a draft of this paper and encouraging us to communicate these results. SP acknowledges support of a DST INSPIRE Faculty Award. SP also thanks the Department of Theoretical Physics (DTP), Tata Institute for Fundamental Research (TIFR), and the International Centre for Theoretical Sciences, TIFR for hospitality where part of this work was performed. 
RS would like to thank Gautam Mandal and Shiraz Minwalla for discussions. RS would also like to thank the students in the DTP Students'
Room at TIFR for all the discussions and help. This work was also partly supported by the Infosys Endowment for the study of the Quantum Structure of Space Time.

\newpage
\section*{Appendix}
\appendix

\section{Calculating the Scalar Spectrum}
Here we present some details for the calculations of the scalar spectrum. The results turn out to be independent of the ratio of $\lambda_2/\lambda_1$. Below, we denote the two-point function in position space as:
\be
G(x,0) = 
-\l^{-1/2}\bigg[\f1{d_{\g}\pi^{d}}\f{\G(3d/4+1/2)}{\G(1/2-d/4)}\bigg]^{1/4}\f{\sl{x}}{(x^2)^{d/4+1/2}} \equiv -\l^{-1/2} \tilde{A} \f{\sl{x}}{(x^2)^{d/4+1/2}}
\ee

\subsection{Integrals and Identities}

Using Equation 2.19 of \cite{Giombi:2017dtl}, we can evaluate most of the integrals that arise in this paper: 
 \begin{equation}
 \int d^d x \frac{(x \cdot z)^s}{x^{2\alpha}(x-y)^{2\beta}}   =  L_{d,s}(\alpha,\beta) \frac{(y\cdot z)^s}{(y^2)^{\alpha+\beta-d/2}} \label{2.19}
 \end{equation}
 Here $z$ is a null polarization vector, satisfying $z^2=0$. 

Using the operator (see e.g., \cite{Dobrev:1975ru, Belitsky:2007j, Costa:2011mg, Giombi:2016hkj})
\begin{equation}
 D^\mu = \partial_{z_\mu} + \frac{1}{d/2-1} z_\nu \partial_{z_\nu} \partial_{z_\mu} -\frac{1}{d-2} z^\mu \partial_{z_\nu}\partial_{z^\nu},
 \end{equation}
 one can translate Equation \eqref{2.19} into the following simple formulas:
 \begin{eqnarray}
  \int d^d x \frac{1}{x^{2\alpha}(x-y)^{2\beta}} &  = & L_{d,0}(\alpha,\beta) \frac{1}{(y^2)^{\alpha+\beta-d/2}} \\
  \int d^d x \frac{x_\mu}{x^{2\alpha}(x-y)^{2\beta}} & = & L_{d,1}(\alpha,\beta)\frac{y_\mu }{(y^2)^{\alpha+\beta-d/2}} \\
  \int d^d x \frac{x_\mu x_\nu }{x^{2\alpha}(x-y)^{2\beta}} & = & \frac{1}{(y^2)^{\alpha+\beta-d/2}} \left( L_{d,2}(\alpha,\beta)y_\mu y_\nu + \frac{y^2 \eta_{\mu\nu}}{d} \left(L_{d,0}(\alpha-1,\beta)-L_{d,2}(\alpha,\beta)\right)   \right) \\
  \int d^d x \frac{x_\mu x_\nu x_\rho}{x^{2\alpha}(x-y)^{2\beta}} & = & \frac{1}{(y^2)^{\alpha+\beta-d/2}} \left( L_{d,3}(\alpha,\beta)y_\mu y_\nu y_\rho + \frac{y^2 \eta_{(\mu\nu}y_{\rho)}}{d+2} \left(L_{d,1}(\alpha-1,\beta)-L_{d,3}(\alpha,\beta)\right)   \right)  \nonumber
 \end{eqnarray}
 where $\eta_{(\mu\nu}y_{\rho)} =\eta_{\mu\nu}y_\rho + \eta_{\mu\rho}y_\nu+\eta_{\nu\rho}y_\mu$.

\subsection{Parity-Odd Scalar Eigenvalue}

\label{parity-odd-calculation}
For the parity odd scalar eigenvalue, the eigenvalue equation can be simplified to take the form:
\begin{equation}
\begin{split}
\frac{g_{\text{odd}}(d,\tau)}{|x_1|^{d/2-\tau}} & = -\lambda^2 \int dx dy ~G(x_1,x) v_\text{odd} (x,y)G(y,0) \text{ Tr} \left(G(x,y)G(y,x)\right) \\ & = (\tilde{A}^4 d_\gamma) \gamma_\mu \gamma_\nu \int dx dy ~ \frac{(x_1-x)^\mu y^\nu (x-y)^2}{|x-y|^{3d/2+2-\tau}|y|^{d/2+1}|x_1-x|^{d/2+1}} \\
& = (\tilde{A}^4 d_\gamma) \gamma_\mu \gamma_\nu \int dx dy ~ \frac{(x_1-x)^\mu y^\nu}{|x-y|^{3d/2-\tau}|y|^{d/2+1}|x_1-x|^{d/2+1}} \\
& = (\tilde{A}^4 d_\gamma) L_{d,1}\left(\frac{d+2}{4},\frac{3d}{4}-\frac{\tau}{2} \right) \gamma_\mu \gamma_\nu \int dx  ~ \frac{(x_1-x)^\mu x^\nu}{|x|^{d+1-\tau}|x_1-x|^{d/2+1}} \\
& = (\tilde{A}^4 d_\gamma) L_{d,1}\left(\frac{d+2}{4},\frac{3d}{4}-\frac{\tau}{2} \right) \left( L_{d,1}\left(\frac{d+1-\tau}{2},\frac{d+2}{4}\right)-L_{d,0}\left(\frac{d-1-\tau}{2},\frac{d+2}{4} \right) \right)\frac{1}{|x_1|^{d/2-\tau}}
\end{split}
\end{equation}
Here we set $x_2=0$, and $$\tilde{A}^4 d_\gamma = \frac{1}{\pi^d}\frac{\Gamma(3d/4+1/2)}{\Gamma(1/2-d/4)}.$$ Note that, in the first line, the integral only depends  on $
\lambda^2=(\lambda_1^2+\lambda_2^2) -2\lambda_1\lambda_2/d_\gamma$, the same quantity which appears in the two-point function, so the spectrum is independent of the ratio between $\lambda_1$ and $\lambda_2$. 

Thus we have
\begin{equation} \begin{split}
g_{\text{odd}}(d,\tau) & = (\tilde{A}^4 d_\gamma) L_{d,1}\left(\frac{d+2}{4},\frac{3d}{4}-\frac{\tau}{2} \right) \left( L_{d,1}\left(\frac{d+1-\tau}{2},\frac{d+2}{4}\right)-L_{d,0}\left(\frac{d-1-\tau}{2},\frac{d+2}{4} \right) \right) \\
 & = -\frac{\Gamma \left(\frac{3 d}{4}+\frac{1}{2}\right) \Gamma
   \left(\frac{d}{4}-\frac{\tau }{2}\right) \Gamma
   \left(\frac{\tau }{2}-\frac{d}{4}\right)}{\Gamma
   \left(\frac{1}{2}-\frac{d}{4}\right) \Gamma \left(\frac{3
   d}{4}-\frac{\tau }{2}\right) \Gamma
   \left(\frac{d}{4}+\frac{\tau }{2}\right)} \\
   & = \frac{4 \cos \left(\frac{\pi 
   d}{4}\right) \Gamma
   \left(\frac{3
   d}{4}+\frac{1}{2}\right)
   \Gamma
   \left(\frac{d+2}{4}\right)
   \csc \left(\frac{1}{4} \pi 
   (d-2 \tau )\right)}{(d-2
   \tau ) \Gamma \left(\frac{3
   d}{4}-\frac{\tau
   }{2}\right) \Gamma
   \left(\frac{1}{4} (d+2 \tau
   )\right)} \end{split}
\end{equation}

For $d=1$, this reduces to:
\begin{equation}
g_{\text{odd}}(1,\tau) = \frac{\tan \left(\frac{1}{4}
   (2 \pi  \tau +\pi
   )\right)}{1-2 \tau }
\end{equation}
which agrees with \cite{Klebanov:2016xxf}.

\subsection{Parity-Even Scalar Eigenvalue}
\label{parity-even-calculation}
Here, the eigenvalue equation can be simplified to take the form:
\begin{eqnarray}
g_{\text{even}}(d,\tau) \frac{\slashed{x}_1}{|x_1|^{d/2-\tau+1}} & = & -3\lambda^2 \int dx dy ~G(x_1,x) v_\text{even} (x,y)G(y,0) \text{ Tr} \left(G(x,y)G(y,x)\right) \\
 	&=& 3(\tilde{A}^4 d_\gamma) \gamma_\mu \gamma_\rho \gamma_\nu \int dx dy \frac{(x_1-x)^\mu (x-y)^\rho y^\nu}{|x-y|^{3d/2+1-\tau}|y|^{d/2+1}|x_1-x|^{d/2+1}} \\
 	&=& 3(\tilde{A}^4 d_\gamma) K_1 \gamma_\mu  \int dx \frac{(x_1-x)^\mu}{|x|^{d-\tau}|x_1-x|^{d/2+1}} \\
 	& = & 3(\tilde{A}^4 d_\gamma) K_1 K_2 \frac{\slashed{x}_1}{|x_1|^{d/2-\tau+1}}
\end{eqnarray}
where 
$$K_1 = L_{d,1}\left( \frac{d+2}{4},\frac{3d}{4}+\frac{1-\tau}{2} \right) -L_{d,0}\left(\frac{d-2}{4},\frac{3d}{4}+\frac{1-\tau}{2} \right),$$ and
$$K_2=L_{d,0}\left(\frac{d-\tau}{2},\frac{d+2}{4} \right)-L_{d,1}\left(\frac{d-\tau}{2},\frac{d+2}{4} \right).$$
 
We have,
\begin{equation}\begin{split}
g_{\text{even}}(d,\tau) &  = 3 (\tilde{A}^4 d_\gamma) K_1 K_2 \\
	&=
	-\frac{3 \cos \left(\frac{\pi  d}{4}\right) \Gamma \left(\frac{3 d}{4}+\frac{1}{2}\right) \Gamma \left(\frac{d+2}{4}\right) \sec \left(\frac{1}{4} \pi 
   (d-2 \tau )\right)}{\Gamma \left(\frac{1}{4} (3 d-2 \tau +2)\right) \Gamma \left(\frac{1}{4} (d+2 \tau +2)\right)}.
   \end{split}
\end{equation}

For $d=1$ this reduces to 
\begin{equation}
g_{\text{even}}(d,\tau) = \frac{3 \cot \left(\frac{1}{4} (2 \pi  \tau +\pi )\right)}{2 \tau -1}
\end{equation}
which agrees with \cite{Klebanov:2016xxf}.

\bibliographystyle{ssg}
\bibliography{tensor}

\end{document}